\documentclass[superscriptaddress,reprint,amsmath,amssymb,aps,pra,floatfix]{revtex4-2}

\usepackage{graphicx}
\usepackage{dcolumn}
\usepackage{bm}
\usepackage{xcolor}
\usepackage[colorlinks = true,
             linkcolor = blue,
             urlcolor  = blue,
             citecolor = blue,
             anchorcolor = blue]{hyperref}

\begin{document}

\title{{Dynamics of one-dimensional Bose-Josephson Junction in a Box Trap:\\ From Coherent Oscillations to Many-Body Dephasing and Dynamical Freezing}}

\author{Abhik Kumar Saha\,\,\href{https://orcid.org/0000-0001-8168-6742}
{\includegraphics[scale=0.05]{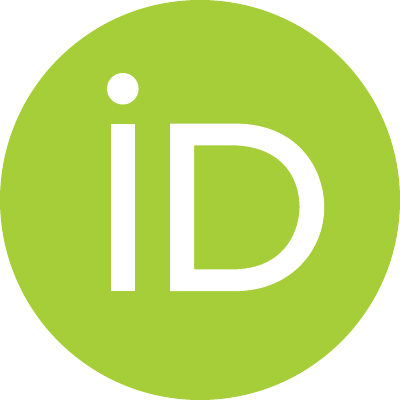}}}
\email{saha.abhikkumar.3k@kyoto-u.ac.jp}
\affiliation{Department of Physics, Kyoto University, Kitashirakawa Oiwakecho, Sakyo-ku, Kyoto 606-8502, Japan}
\affiliation{School of Physical Sciences, Indian Association for the Cultivation of Science, Jadavpur, Kolkata 700032, India}
\author{L. F. Calazans de Brito\,\,\href{https://orcid.org/0000-0001-6889-0810}
{\includegraphics[scale=0.05]{orcidid.pdf}}}
\affiliation{Instituto de Física, Universidade de São Paulo, CEP 05508-090, SP, Brazil.}
\author{Cesare Vianello\,\,\href{https://orcid.org/0009-0001-1136-5924}
{\includegraphics[scale=0.05]{orcidid.pdf}}}
\affiliation{Dipartimento di Fisica e Astronomia ``Galileo Galilei'', Università di Padova, Via Marzolo 8, 35131 Padua, Italy}
\affiliation{INFN Sezione di Padova, Via Marzolo 8, 35131 Padua, Italy}
\author{Rhombik Roy\,\,\href{https://orcid.org/0000-0002-7546-3320}
{\includegraphics[scale=0.05]{orcidid.pdf}}}
\affiliation{Department of Physics, University of Haifa, Haifa 3498838, Israel.}
\affiliation{Haifa Research Center for Theoretical Physics and Astrophysics,
University of Haifa, Haifa 3498838, Israel.}
\author{Romain Dubessy\,\,\href{https://orcid.org/0000-0002-9448-2535}
{\includegraphics[scale=0.05]{orcidid.pdf}}}
\affiliation{Aix-Marseille University, CNRS UMR 7345, PIIM, 13397, Marseille, France}
\author{Barnali Chakrabarti\,\,\href{https://orcid.org/0000-0002-6320-9894}
{\includegraphics[scale=0.05]{orcidid.pdf}}}
\email{barnali@if.usp.br}
\affiliation{Instituto de F\'{\i}sica de S\~{a}o Carlos, Universidade de S\~{a}o Paulo, 
CP 369, 13560-970 S\~{a}o Carlos, SP, Brazil}
\affiliation{ Laboratório de Física Teórica e Computacional, Departamento de Física,
Universidade Federal de Pernambuco, 50670-901 Recife, Pernambuco, Brazil}
\author{Arnaldo Gammal\,\,\href{https://orcid.org/0000-0003-4720-3203}
{\includegraphics[scale=0.05]{orcidid.pdf}}}
\affiliation{Instituto de Física, Universidade de São Paulo, CEP 05508-090, SP, Brazil.}

\date{\today}
\begin{abstract}

Understanding how coherent quantum dynamics give way to correlation-dominated behavior in low-dimensional systems remains a central challenge in quantum many-body physics. Here, we investigate a one-dimensional Bose-Josephson junction confined in a box trap using the multiconfigurational time-dependent Hartree method for bosons (MCTDHB). By varying the interaction strength and initial population imbalance, we identify distinct dynamical regimes governed by the competition between coherence and correlation-induced fragmentation. Weak interactions support coherent Josephson oscillations, whereas increasing imbalance leads to damping. At intermediate interaction strength, varying only the initial imbalance induces a crossover from nearly pure coherent oscillations to many-body dephasing with collapse-and-revival dynamics, and ultimately to equilibration accompanied by strong fragmentation and the saturation of many-body observables. In the strongly interacting regime, the system enters a dynamical freezing regime characterized by pronounced fragmentation, well-separated particle-resolved density peaks, and strongly suppressed tunneling. A systematic comparison with the Bose-Hubbard model reveals excellent agreement in the weakly interacting regime, while progressively larger deviations emerge as higher-orbital occupations beyond the two-mode approximation become significant. These results provide a unified picture of the emergence and competition of coherence, many-body dephasing, equilibration, and dynamical freezing, while delineating the regime of validity of the Bose-Hubbard description.

\end{abstract}

\maketitle
\clearpage

\section{INTRODUCTION}

Understanding non-equilibrium dynamics in low-dimensional quantum systems remains a central challenge in modern many-body physics, as reduced dimensionality enhances quantum fluctuations and correlations~\cite{Bloch2008,Lewenstein2007}. Ultracold atomic gases provide an exceptional platform to explore these effects, offering precise control over interactions~\cite{Chin2010,Mies2000}, trapping geometries~\cite{Dubessy_2025, Levy:2007, Gati2007}, and population imbalances. These capabilities have enabled the experimental realization of paradigmatic systems where the interplay of coherence, tunneling, and correlations can be studied in real time~\cite{Betz:2011,Pigneur2018,Hofferberth2007,Qiao2023,Sakmann2009,Sakmann2014Universality,STEFANATOS20192370,Meinert_2014}.

Among such systems, two weakly coupled Bose–Einstein condensates (BECs)—realizing an atomic Bose–Josephson junction (BJJ)—have emerged as a versatile platform for investigating tunneling dynamics~\cite{shenoy:1999,josephson:exp,josephson:exp1,Pigneur2018,josephson:exp2,josephson:exp3,josephson:exp4,Milburn:1997,Zapata:1998,Raghavan:1999,Abad:2015,Levy:2007,Smerzi:2003} as well as many-body correlations~\cite{Schurer:2016,Haldar:2019,Saha2020,Saha:2023,Bhowmik:2020,Boukobza:2010,Betz:2011}. Although the theory of the Josephson junction was originally developed in the context of superconductivity~\cite{josephson:1962,Josephson:proof,SJJ:app}, it can be directly extended to describe two weakly coupled BECs~\cite{shenoy:1999,Raghavan:1999,Milburn:1997} as well as their counterparts with fermionic superfluid atomic samples \cite{josephson:fermi,Josephson-fermi-2:2020,Zaccanti:2019,Burchianti:2018,Xhani:2020,Kwon:2020}. 

In the weakly interacting regime, Josephson oscillations between the two condensates are well captured by mean-field approaches~\cite{Saha2019,shenoy:1999,Xhani2020dynamical,Saha2021,Singh:2020,Polo:2019}, which typically predict nondissipative dynamics. However, recent studies have reported the emergence of dissipation in BJJs~\cite{Mennemann:2021,Yuri:2021,Xhani:2022,Xhani2020dynamical}, particularly when the system is coupled to external bosonic~ baths~\cite{Saha2020,Saha:2023}. For further developments in both theoretical and experimental investigations of dissipative BJJs, see Refs~\cite{Ji:2022,Burchianti:2017,josephson:exp4,Pigneur2018,Bidasyuk:2016,Mennemann:2021,Yuri:2021,Xhani:2022,Xhani2020dynamical}. Moreover, in one-dimensional (1D) systems, interactions and quantum correlations give rise to phenomena such as dephasing, fragmentation~\cite{Mueller2006Fragmentation}, and equilibration, which lie beyond the scope of simple two-mode models~\cite{shenoy:1999} or conventional mean-field descriptions~\cite{Xhani2020dynamical,Saha2021,Singh:2020,Polo:2019}.

When investigating Bose–Josephson junctions, the system geometry plays a central role in determining the underlying dynamics. Most studies have focused on double-well potentials, where two elongated condensates are arranged side by side~\cite{Mennemann:2021,Shin:2004,Schumm:2005}. Alternatively, ring geometries~\cite{Didier:2009,Polo:2019,Koon:arxiv}, in which the condensates are coupled in a head-to-tail configuration, have also been considered. The former is particularly suited for matter-wave interferometry~\cite{Schumm:2005,Gati:2006}, whereas the latter enables the realization of atomtronic circuits~\cite{Amico:2017,Ryu:2020,Edwards:2013}.

In this context, box-shaped potentials~\cite{Murtadho:2025,Saha2021,Tajik:2019} introduce qualitatively distinct features. The flat-bottom confinement combined with hard-wall boundaries leads to discrete and nearly evenly spaced modes, as well as pronounced interference effects. These characteristics significantly modify collective excitations and the dynamics of coherence. Consequently, box traps~\cite{Murtadho:2025,Tajik:2019,van_Es:2010,Zhang:2008} provide a natural platform to investigate the crossover from coherent tunneling to strongly correlated many-body dynamics, including phenomena such as fragmentation, many-body dephasing, ergodic-like relaxation, and interaction-induced dynamical freezing, which are characteristic of low-dimensional strongly interacting systems~\cite{Bloch2008, Girardeau1960, Kinoshita2004, Alon:2008,Streltsov:2007}.

While the Bose-Hubbard (BH) model has become the standard theoretical framework for describing Bose-Josephson junctions~\cite{Cirac_1998,Mazzarella_2011,Vianello_2025,Anglin_2001,Meystre_2005,Tonel_2005,Vianello_2025b}, its validity relies on the assumption that the dynamics are well captured within a restricted two-mode basis~\cite{Zhou2003Josephson,Links2006BoseHubbard,Simon2012Josephson}. This approximation is expected to become increasingly inadequate as interactions strengthen, higher orbitals become populated, and many-body correlations and fragmentation develop. Although the BH model has been extensively employed to study Josephson dynamics, a systematic assessment of its accuracy across different interaction regimes in low-dimensional box-confined systems remains lacking. Such a benchmark is particularly important in view of recent experiments~\cite{Navon2021} probing strongly correlated non-equilibrium dynamics beyond the two-mode regime.

In this work, we present a systematic theoretical study of a 1D BJJ in a box-shaped potential using the multiconfigurational time-dependent Hartree method for bosons (MCTDHB)~\cite{Alon:2007,Streltsov:2007,Alon:2008,Streltsov:2006,Lode:2016} implemented in the MCTDH-X software~\cite{Alon:2007,Lode:2016,lin:2020,MCTDHX,Fasshauer:2016,Lode:2020}, which has been successfully applied to a wide range of related many-body problems in recent years~\cite{Sakmann2009,Sakmann2014Universality,rhombik_acc,rhombik_epjd,rhombik_jcp,rotation_BJJ}. By performing a direct comparison with the BH model, we establish the regime of validity of the effective two-mode description and identify the many-body mechanisms responsible for its breakdown. By varying both the interaction strength and the initial population imbalance, we investigate how interactions, initial conditions, and confinement jointly govern the nonequilibrium dynamics of strongly correlated 1D Bose-Josephson junctions.

Specifically, we focus on: (i) Characterizing the different dynamical regimes, including coherent Josephson oscillations, many-body dephasing, equilibration, and dynamical freezing associated with strong fragmentation; (ii) Investigating the role of the initial population imbalance in governing these dynamical regimes; (iii) Quantifying coherence and correlations through experimentally accessible observables such as the population imbalance and oscillation frequency, while characterizing fragmentation using many-body measures including natural orbital occupations, orbital entropy, and participation ratio; and (iv) Benchmarking the BH model against the MCTDHB results to establish the regime of validity of the two-mode approximation and identify the onset of its breakdown as additional natural orbitals become significantly occupied. To this end, we consider a system of $N$ bosons confined in a 1D hard-wall box, separated into two weakly coupled regions, and initiate the dynamics via a quench of the external potential while keeping the interaction strength fixed.

Our study reveals distinct dynamical regimes in the 1D BJJ, governed by the interplay of interactions and initial population imbalance. Weak interactions give rise to coherent and damped Josephson oscillations, while intermediate interactions exhibit a crossover from coherent oscillations to many-body dephasing and equilibration as the initial imbalance increases. In the strongly interacting regime, the system enters a dynamical freezing regime associated with strong fragmentation, where tunneling is strongly suppressed. Systematic benchmarking against the BH model demonstrates that the two-mode description accurately captures the weakly interacting dynamics but progressively breaks down with increasing interaction strength as multiorbital correlations become significant. These results provide a unified many-body framework for understanding non-equilibrium Josephson dynamics and establish the regime of validity of the BH description for 1D box-confined Bose–Josephson junctions.

The paper is organized as follows. In Sec.~\ref{sec:2}, we present the theoretical framework, including the many-body Hamiltonian, the quench protocol, the MCTDHB and BH methodology, and the observables used to characterize the system. In Sec.~\ref{sec:3}, we discuss the results, divided into three subsections corresponding to the different dynamical regimes. Finally, Sec.~\ref{sec:4} summarizes our main findings and conclusions.

\section{Theoretical framework}\label{sec:2}

\subsection{Hamiltonian and quench protocol}

The many-body Hamiltonian for $N$ bosons confined in a 1D trap potential $V_{\rm trap}(x)$ is given by
\begin{equation}
\begin{aligned}
\hat{H} &=
\sum_{i=1}^{N}
\left[
-\frac{\hbar^{2}}{2m}\frac{d^{2}}{dx_i^{2}}
+V_{\mathrm{trap}}(x_i)
\right]
+\sum_{i<j}^{N}W(x_i-x_j)\\
&=
\sum_{i=1}^{N}h(x_i)
+\sum_{i<j}^{N}W(x_i-x_j).
\end{aligned}
\label{eq:Hamiltonian}
\end{equation}
Here, $x_i$ denotes the coordinate of the $i$-th particle. The external trapping potential is composed of three Gaussian barriers,
\begin{equation}
\begin{aligned}
V_{\rm ext}(x) &= V_1 \exp\left[-\frac{x^2}{\sigma^2}\right]
+ V_0 \exp\left[-\frac{(x-L)^2}{\sigma^2}\right] \\
&\quad + V_0 \exp\left[-\frac{(x+L)^2}{\sigma^2}\right].
\end{aligned}
\end{equation}

\begin{figure}[htbp]
\centering
\includegraphics[width=0.40\textwidth]{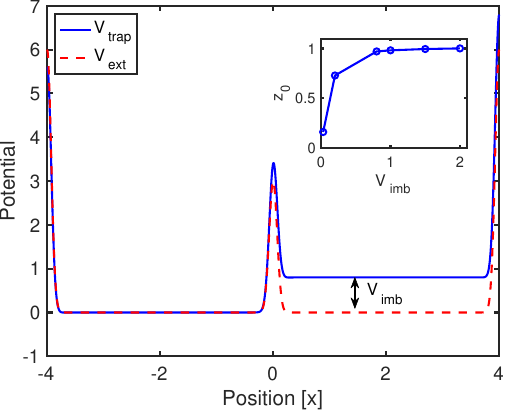}
\caption{Setup and quench protocol considered in the present work. Initially, a 1D Bose gas is prepared in two reservoirs with a finite population imbalance, induced by an offset potential $V_{\rm imb}=0.8$ (indicated by the double arrow), superimposed on the trapping potential with central barrier height $V_1=3.0$, as illustrated by the solid blue curve corresponding to the trapping potential $V_{\rm trap}$. At $t=0$, the offset is suddenly removed, resulting in a modified potential landscape $V_{\rm ext}$ shown by the red dashed curve. The inset displays the initial population imbalance $z_{0}$
as a function of  $V_{\rm imb}$ obtained using the MCTDHB method.}
\label{Fig1}
\end{figure}

The barriers are located at $x=0$ and $x=\pm L$, with the outer Gaussian barriers providing an effective implementation of periodic boundary conditions. For sufficiently large $V_0$, the resulting potential generates a box-like confinement with a tunable central barrier of strength $V_1$, thereby forming two weakly coupled reservoirs, as shown in Fig.~\ref{Fig1}. Throughout this work, we use dimensionless units (see Appendix~\ref{appendix:c}), fix the Gaussian width to $\sigma=0.1$, and consider a box of total length $2L=8$ (i.e., $x\in[-4,4]$), while varying the central barrier strength within the range $V_1\in[0.5,4.5]$.

To generate an initial population imbalance between the two reservoirs, we add an auxiliary offset potential
\begin{equation}
V_{\rm offset}(x)=\frac{V_{\rm
imb}}{2}\left[\tanh{\left(\frac{x}{\sigma}\right)}+\tanh{\left(\frac{L-x}{\sigma}\right)}\right],
\end{equation}
where $V_{\rm imb}$ controls the magnitude of the imbalance, as shown in the inset of Fig.~\ref{Fig1}. The system is initialized in the ground state of the combined potential, $V_{\rm trap}(x)$, described by the blue solid line in Fig.~\ref{Fig1},
\begin{equation}
V_{\rm trap}(x)=V_{\rm ext}(x)+V_{\rm offset}(x),
\end{equation}
obtained via imaginary-time propagation. At $t=0$, the offset potential is suddenly removed, and the system evolves in real time under $V_{\rm ext}(x)$ alone.

The bosons interact via a contact potential
\begin{equation}
W(x_i-x_j) = \Lambda\delta(x_i-x_j),
\end{equation}
with tunable interaction strength $\Lambda$. The non-equilibrium dynamics of the system are governed by the time-dependent many-body Schrödinger equation
\begin{equation}
\hat{H} \vert\Psi(t) \rangle = i \hbar \frac{\partial}{\partial t}\vert\Psi(t)\rangle.
\end{equation}

\subsection{Many-body methods}

To investigate the non-equilibrium dynamics of the 1D BJJ, we employ two complementary many-body approaches. The primary framework is the MCTDHB, which provides a numerically exact treatment of correlation effects within a systematically improvable basis of time-dependent orbitals. To assess the validity of reduced descriptions and quantify the role of multiorbital correlations, we additionally employ the two-site BH model, which is based on a fixed two-mode approximation of the many-body Hilbert space. A direct comparison between these two approaches enables us to establish the regime of validity of the effective two-mode description and to identify the many-body mechanisms responsible for its breakdown as interactions and fragmentation become increasingly important. For completeness, selected results are also compared with the corresponding semi-classical (mean-field) description, whose formulation is summarized in Appendix~\ref{appendix:semiclassical}.

\subsubsection{Multiconfigurational time-dependent Hartree method for bosons (MCTDHB)}

MCTDHB is a first-principles many-body approach based on a time-dependent variational principle, in which the many-body wavefunction is expanded in a set of $M$ self-consistently optimized single-particle orbitals. Both the expansion coefficients and the orbitals are determined dynamically, enabling a systematically improvable description of correlation and fragmentation beyond mean-field theory.

Because the single-particle basis evolves in time, MCTDHB is particularly well suited for strongly nonequilibrium settings where the development of correlations and occupation of higher natural orbitals play a central role. The method has been extensively applied to ground-state and dynamical properties of ultracold bosonic systems, including strongly correlated regimes~\cite{fischer:2015,Bera:2019,rhombik_EPJPLUS,chatterjee:2018,chatterjee:2019,chatterjee:2020,Hughes:2023}.

The many-body wavefunction is written as a linear combination of time-dependent permanents:
\begin{equation}
\vert \Psi(t)\rangle = \sum_{\vec{n}}^{} C_{\vec{n}}(t)\vert \vec{n};t\rangle,
\label{many_body_wf}
\end{equation}
where $\vec{n} = (n_1,n_2, \dots ,n_M)$ denotes the occupation of the orbitals, satisfying $n_1 + n_2 + \dots +n_M = N$. Each permanent is constructed as
\begin{equation}
\vert \vec{n};t\rangle = \prod^M_{k=1}\left[ \frac{(\hat{b}_k^\dagger(t))^{n_k}}{\sqrt{n_k!}}\right] |0\rangle,
\label{many_body_wf_2}
\end{equation}
with $|0\rangle$ the vacuum state and $\hat{b}_k^\dagger(t)$ creating a particle in the $k$-time-dependent orbital. Both the coefficients 
$C_{\vec{n}}(t)$ and the orbitals $\psi_{k}(x,t)$ are variationally optimized at each time step, allowing the method to dynamically track correlations arising from interactions.

The time-adaptive basis in MCTDHB allows the wavefunction to dynamically adjust to the evolving many-body correlations, ensuring that the most relevant orbitals are always included. The method becomes exact in the limit $M \rightarrow \infty$; in practice, a finite number of orbitals is sufficient when additional orbitals remain negligibly populated and key observables converge. Convergence is checked by systematically increasing $M$ until quantities such as natural orbital occupations, and many-body entropy become insensitive to further increases.

Importantly, by allowing multiple orbitals to be significantly occupied, MCTDHB naturally captures fragmentation, in which the condensate is distributed over more than one orbital. This is essential for describing dynamics beyond mean-field theory, including dephasing, collapse-and-revival phenomena, equilibration, and fragmentation-induced freezing effects in strongly interacting or imbalanced systems. The variational optimization ensures that both the orbitals and their occupation numbers evolve self-consistently~\cite{TDVM81,variational1,variational3,variational4}, providing an accurate representation of the full correlated many-body state throughout the time evolution.

\subsubsection{Two-mode Bose–Hubbard model}
The BH model provides a widely used two-mode description of bosonic Josephson junctions in double-well-like configurations. In this approach, the many-body field operator is restricted to the lowest pair of localized single-particle modes associated with the left and right wells,
\begin{equation}
\hat{\Psi}(x)=
\hat{b}_L \phi_L(x)+\hat{b}_R \phi_R(x),
\label{eq:tm}
\end{equation}
where $\phi_L(x)$ and $\phi_R(x)$ are time-independent localized orbitals. This truncation reduces the full continuum problem to an effective two-mode system.

In second quantization, the Hamiltonian~(\ref{eq:Hamiltonian}) of the many-boson system can be written in terms of the bosonic field operators as
\begin{equation}
\begin{split}
\hat{H}={}&
\int dx\, \hat{\Psi}^{\dagger}(x)\,h(x)\,\hat{\Psi}(x)\\
&+\frac{1}{2}\int dx\,dx'\,
\hat{\Psi}^{\dagger}(x)\hat{\Psi}^{\dagger}(x')
W(x-x')
\hat{\Psi}(x')\hat{\Psi}(x),
\end{split}
\label{Hmbody}
\end{equation}
where $\hat{\Psi}^{\dagger}(x)$ and $\hat{\Psi}(x)$ are the bosonic creation and annihilation field operators, respectively, and $h(x)$ and $W(x-x')$ denote the one-body and two-body operators introduced above.

Substituting Eq.~\eqref{eq:tm} into the many-body Hamiltonian~\eqref{Hmbody}, and retaining only on-site interaction terms and nearest-neighbor tunneling processes, leads to the BH Hamiltonian,
\begin{equation}
\hat{H}_{\mathrm{BH}} =
- J\!\left(\hat{b}_L^\dagger \hat{b}_R+\hat{b}_R^\dagger \hat{b}_L\right) + \frac{U}{2}\!\left(
\hat{b}_L^{\dagger}\hat{b}_L^{\dagger}\hat{b}_L\hat{b}_L
+\hat{b}_R^{\dagger}\hat{b}_R^{\dagger}\hat{b}_R\hat{b}_R
\right),
\end{equation}
where $J$ denotes the tunneling amplitude and $U$ is the effective interaction strength.

The even-symmetry ground state $\phi_g$ and the odd-symmetry first excited state $\phi_u$ of the single-particle box-trap Hamiltonian $h(x)$ allow us to construct localized left- and right-mode functions as
\begin{equation}
\phi_{L,R}(x)=\frac{1}{\sqrt{2}}\left(\phi_g(x)\pm\phi_u(x)\right).
\end{equation}
The tunneling parameter in the resulting two-mode description is then defined as
\begin{equation}
J=-\int dx\,\phi_L^*(x)h(x)\phi_R(x).
\end{equation}
The on-site interaction parameter is given by
\begin{equation}
U=\Lambda\int dx\,|\phi_{L,R}(x)|^4.
\end{equation}

The many-body state within this two-mode approximation is expanded in the occupation-number basis as
\begin{equation}
|\Psi_{\mathrm{BH}}(t)\rangle =
\sum_{n_L=0}^{N} C_{n_L}(t)\,|n_L, N-n_L\rangle,
\end{equation}
and its time evolution is governed by the Schr\"odinger equation
\begin{equation}
\hat{H}_{\mathrm{BH}} |\Psi_{\mathrm{BH}}(t)\rangle=i\hbar \frac{d}{dt} |\Psi_{\mathrm{BH}}(t)\rangle.
\end{equation}
For a quantitative comparison, we determine the BH parameters $J$ and $U$ from the localized single-particle modes for each interaction strength and barrier height considered.

The essential difference between the two approaches lies in the treatment of the single-particle basis. The BH model employs a fixed two-mode basis, restricting the dynamics to a predefined subspace of the many-body Hilbert space. In contrast, MCTDHB uses a variationally optimized, time-dependent orbital basis that adapts during the evolution, enabling the dynamical build-up of correlations, fragmentation, and the occupation of additional effective modes beyond the two-mode approximation. As a consequence, the BH model is expected to provide a reliable description when the dynamics is dominated by two-mode tunneling, whereas MCTDHB remains applicable across all interaction strengths considered here by accounting for the dynamical emergence of multiorbital correlations and fragmentation. To quantify the crossover between these descriptions, we perform a systematic comparison throughout the Sec.~\ref{sec:3}. This benchmark establishes the range of validity of the BH model and identifies the interaction regimes in which multiorbital effects become essential for describing the non-equilibrium dynamics.

\subsection{Observables}
To characterize the non-equilibrium dynamics, we monitor several experimentally and theoretically relevant observables:

{\textit{ Population Imbalance}} $z(t)$:
\begin{equation}
    z(t)= \frac{N_{L}(t) - N_{R}(t)} {N}
\end{equation}
where $N_{L}(t)$  and $N_{R}(t)$ are the instantaneous particle numbers in the left and right wells, respectively. This quantifies the tunneling dynamics and Josephson oscillations between the two reservoirs.

{\textit{ Oscillation frequency}} $\omega$: Obtained from the Fourier spectrum of $z(t)$. To extract the frequency, we calculate the power spectrum density of $z(t)$, defined as $|z(\omega)|^2$, where $z(\omega)$ is the Fourier transform of $z(t)$ providing insight into coherent tunneling and interaction-induced shifts. We always use normalized power spectrum density, obtained by dividing $|z(\omega)|^2$ by its maximum value.  

{\textit{One-body reduced density matrix and Fragmentation}}: The reduced one-body density matrix is defined as
\begin{equation}
\rho^{(1)}(x, x^{\prime}; t) = \left< \Psi(t) \right| \hat{\Psi}^{\dagger}(x^{\prime}) \hat{\Psi}(x) \left| \Psi(t) \right>.
\end{equation}
Fragmentation is the hallmark of MCTDHB when more than one single particle state becomes significantly occupied. The dynamics of occupation in different orbitals offers a measure for the dynamical fragmentation in the non-equilibrium dynamics of the quantum quench.

We define fragmentation from the natural occupations $n_i$, i.e. the population of the natural orbitals, which are the eigenvalues of the reduced one-body density matrix, i.e, 
\begin{equation}
\rho^{(1)}(x, x'; t) = \sum_j {n_j(t)} \phi_j(x,t) \phi_j^*(x^{\prime},t) 
\label{eq:rho1}
\end{equation}
as a spectral decomposition. $\phi_j(x,t)$ are the natural orbitals and $n_j(t)$ is the occupation in the respective natural orbitals, at time $t$. In our indexing convention, the natural orbitals are ranked in order of decreasing occupation, i.e. $n_1 \ge n_2 \ge .....  \ge n_M $.

{\textit{ Coefficient entropy}} $S_C(t)$:  Calculated from the expansion coefficients $C_{\vec{n}}(t)$ as
\begin{equation}
 S_{C}(t) = -\sum_{\vec{n}} | C_{\vec{n}}(t)|^2 \ln |C_{\vec{n}}(t)|^{2},
\end{equation}
capturing the spreading of the many-body wavefunction across different configurations.

{\textit{ Orbital entropy}} $S_n(t)$:  Computed from the eigenvalues of the reduced one-body density matrix,
\begin{equation}
S_n(t)= - \sum_{i} n_i(t) \ln [n_i(t)],
\end{equation}
quantifying the degree of fragmentation and correlation among the orbitals.

{\textit{Participation ratio (PR)}}: Defined from the expansion coefficients as
\begin{equation}
    \textrm{PR}(t)= \frac{1} {\sum_{\vec{n}}| C_{\vec{n}}(t)|^4},
\end{equation}
measuring the effective number of many-body configurations contributing to the wavefunction and serving as an indicator of correlations.

\section{Results and discussions}\label{sec:3}

We begin by preparing the system in the ground state of a 1D box potential with an imposed potential offset that creates a finite population imbalance between the two sides of the junction. The presence of the central barrier partitions the system into two weakly coupled regions, while the offset controls the initial occupation asymmetry. Throughout this work, we focus on a system of $N=10$ bosons. The influence of particle number is examined separately in Appendix~\ref{appendix:finitesize}, where we present a finite-size scaling analysis using the standard mean-field interaction scaling protocol. 

The interplay of interaction strength and barrier height determines the degree of coherence and the emergence of many-body correlations already in this initial state, with signatures of partial fragmentation becoming increasingly pronounced in the strongly interacting regime. This imbalanced configuration provides a well-defined starting point for the subsequent nonequilibrium dynamics. At time $t=0$, the offset potential is suddenly removed, thereby quenching the system out of equilibrium and initiating the dynamical evolution.

To systematically characterize the ensuing dynamics, we monitor a set of observables including the population imbalance $z(t)$, oscillation frequency, natural orbital occupations, orbital entropy, coefficient entropy, participation ratio, and correlation functions.

We compare the full many-body MCTDHB results with both the BH model and the corresponding semi-classical Josephson equations. In the weakly interacting regime, the three approaches show consistent short-time dynamics, whereas systematic deviations emerge with increasing interaction strength and growing occupation of higher natural orbitals. In particular, the onset of multiorbital correlations leads to a gradual breakdown of the two-mode BH description, while the semi-classical model fails earlier due to the absence of quantum fluctuations and correlation effects.

As we vary the interaction strength and initial conditions, three distinct dynamical regimes emerge: (i) weakly interacting Josephson oscillations, characterized by coherent or weakly damped tunneling between the two regions; (ii) intermediate interactions with imbalance-driven dephasing and equilibration, where many-body correlations induce collapse-and-revival dynamics and eventual equilibration for sufficiently large initial imbalances; and (iii) strongly interacting dynamics, where the system enters a frozen regime associated with strong fragmentation and suppressed tunneling.

In the following subsections, we analyze each of these regimes in detail, focusing on the role of interactions and initial population imbalance in shaping coherence, fragmentation, and relaxation dynamics across different theoretical descriptions.

To characterize the interaction regimes discussed in the following subsections, we use the dimensionless Lieb-Liniger parameter $\gamma$ together with the dimensionless healing-length parameter $\zeta$. These quantities provide complementary measures of the interaction strength and the healing length of the trapped system, and serve as the primary indicators for distinguishing weakly, intermediate, and strongly interacting regimes.

In the MCTDHB simulations, we work in dimensionless units with $\hbar=m=1$. The microscopic interaction strength is characterized by the 1D coupling constant $g_{1\mathrm{D}}$. The relevant quantities used to characterize the interaction regime are defined as $g_{1\mathrm{D}}=\Lambda$, $n_{1\mathrm{D}}=N/2L$, $\gamma=g_{1\mathrm{D}}/n_{1\mathrm{D}}$, $\xi=1/\sqrt{2g_{1\mathrm{D}}n_{1\mathrm{D}}}$, and $\zeta=2L/\xi$.

Here, $n_{1\mathrm{D}}$ denotes the mean linear density, $\xi$ is the healing length, and $2L$ is the system size. While the dimensionless Lieb-Liniger parameter $\gamma$ quantifies the interaction strength relative to the 1D density and $\zeta$ compares the system size to the healing length. These quantities are used throughout this work to characterize the crossover from coherent Josephson dynamics to strongly correlated many-body behavior.

\subsection{Weak-interaction regime: coherent and damped Josephson oscillations}

We first consider the weakly interacting regime characterized by $\Lambda=0.01$ for $N=10$ bosons. The corresponding interaction parameters are $\gamma \approx 8\times10^{-3}$ and $\zeta \approx 1.265$, placing the system deep in the weakly interacting regime. The small value of $\gamma$ indicates negligible interaction-induced correlations, while $\zeta \simeq 1$ implies that the healing length is comparable to the size of the junction, reflecting weak interactions for which kinetic energy dominates the density profile. In this regime, the dynamics is dominated by coherent tunneling between the two weakly coupled regions, leading to Josephson-like oscillations of the population imbalance $z(t)$. The small value of $\gamma$ implies that interaction-induced depletion and fragmentation remain weak, and the system is therefore close to the mean-field coherent limit.

All three descriptions—MCTDHB, the BH model, and the semi-classical Josephson equations—reproduce the short-time oscillatory dynamics consistently in this regime. However, weak damping of the oscillations is observed beyond mean-field predictions, originating from many-body dephasing effects captured by the BH model and fully resolved within MCTDHB.

\begin{figure}[t]
\centering
\includegraphics[width=0.45\textwidth]{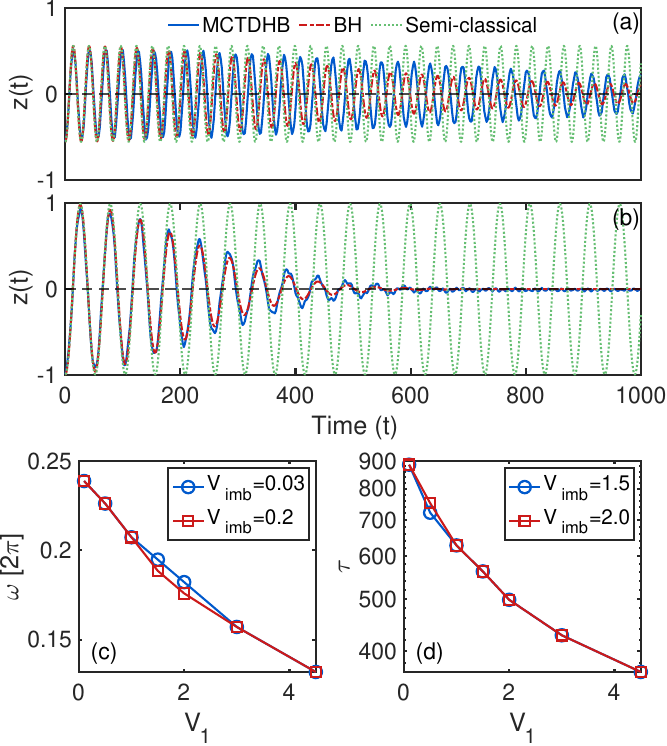}
\caption{Dynamics of the population imbalance $z(t)$ in the weak-interaction regime ($\Lambda=0.01$). (a) Small initial imbalance ($V_{\mathrm{imb}}=0.2$) and small barrier height ($V_1=0.5$): all three approaches—semi-classical Josephson equations, BH model, and MCTDHB—exhibit coherent Josephson oscillations, with only very weak damping appearing in BH and MCTDHB at long times. (b) Larger initial imbalance ($V_{\mathrm{imb}}=1.0$) and higher barrier ($V_1=4.5$): the semi-classical model shows undamped oscillations, whereas BH and MCTDHB display pronounced damping and equilibration of $z(t)$ due to many-body dephasing. (c) Variation of the oscillation frequency $\omega$ obtained from MCTDHB simulations as a function of the central barrier height $V_1$ for two different initial population imbalances, $V_{\rm imb}=0.03$ and $V_{\rm imb}=0.2$. (d) Dependence of the decay time $\tau$ (in log scale) obtained from MCTDHB simulations on the barrier height $V_1$ for two representative large-imbalance cases, showing an approximately exponential scaling and indicating strong barrier-controlled damping and equilibration dynamics.
}
\label{Fig2}
\end{figure}

Fig.~\ref{Fig2}(a) shows the time evolution of the population imbalance $z(t)$ for small initial imbalance ($V_{\mathrm{imb}}=0.2$) and small barrier height ($V_1=0.5$). The corresponding BH parameters are $U=2.9062\times 10^{-3}$ and $J=1.0839\times 10^{-1}$. In this regime, the system exhibits nearly coherent Josephson oscillations over the entire simulation time. The semiclassical Josephson equations predict perfectly undamped oscillations, as expected from the absence of quantum fluctuations. In contrast, both the BH model and MCTDHB show only very weak damping of the oscillations, with a small reduction in amplitude emerging at long times. It is important to observe that the BH model slightly overestimates the Josephson frequency. Consequently, over many oscillation periods the BH and MCTDHB solutions gradually accumulate a phase difference, reaching approximately $\pi$ around $t\simeq 500$ and about $2\pi$ around $t\simeq 1000$.

Fig.~\ref{Fig2}(b) presents the dynamics for larger initial imbalance ($V_{\mathrm{imb}}=1.0$) and higher barrier height ($V_1=4.5$). The corresponding BH parameters are $U=3.3004\times 10^{-3}$ and $J=5.9553\times 10^{-2}$. In this regime, the semi-classical model again predicts persistent undamped oscillations. However, both BH and MCTDHB exhibit a clear damping of Josephson oscillations, with $z(t)$ gradually decaying and approaching an equilibrium-like value near zero on the same characteristic timescale. This behavior reflects the emergence of many-body dephasing and correlation-induced relaxation, which is absent at the mean-field level but captured within both BH and fully resolved in MCTDHB. The enhanced damping with increasing initial imbalance highlights the strong role of interaction-driven dephasing in the breakdown of simple Josephson dynamics. The improved agreement between BH and MCTDHB in Fig.~\ref{Fig2}(b) in contrast with Fig.~\ref{Fig2}(a) is mainly due to the higher barrier, which enhances the localization of the left- and right-well orbitals and thereby increases the accuracy of the BH description.

\begin{figure}[t]
\centering
\includegraphics[width=0.45\textwidth]{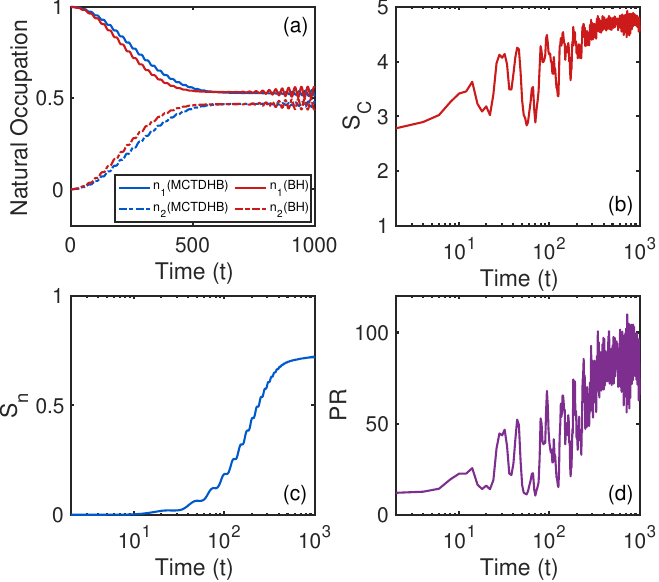}
\caption{Same parameters as in Fig.~\ref{Fig2}(b). (a) Time evolution of the occupations of the first two natural orbitals obtained from MCTDHB (blue-lines) and the BH model (red-lines), showing excellent agreement throughout the dynamics and the emergence of two-fold fragmentation. (b) Coefficient entropy $S_C(t)$, (c) orbital entropy $S_n(t)$, and (d) participation ratio (PR) as functions of time. The orbital entropy rapidly approaches a quasi-stationary value associated with two-fold fragmentation, while $S_C(t)$ and the participation ratio evolve toward long-time quasi-stationary behavior with finite temporal fluctuations characteristic of an isolated finite quantum system.}
\label{Fig3}
\end{figure}

To further quantify the tunneling dynamics, we employ MCTDHB simulations and extract the Josephson oscillation frequency as a function of the central barrier height $V_1$ for two representative initial population imbalances, $V_{\mathrm{imb}}=0.03$ and $V_{\mathrm{imb}}=0.2$. As shown in Fig.~\ref{Fig2}(c), the oscillation frequency decreases monotonically with increasing barrier height, reflecting the progressive suppression of tunneling through the central barrier. The dependence is similar for both initial imbalances, indicating that, in the weak-interaction regime, the tunneling dynamics is primarily controlled by the barrier height. These results confirm that weak interactions preserve global phase coherence and support long-lived Josephson oscillations.

To characterize the damping of Josephson oscillations, we employ MCTDHB simulations and define a decay time $\tau$ as the instant at which $|z(t)|$ drops below one fourth of its initial value. The dependence of $\tau$ on the barrier height $V_1$ for two representative large-imbalance cases is shown in Fig.~\ref{Fig2}(d), exhibiting an approximately exponential scaling. This is a further indication that the damping and equilibration dynamics are strongly controlled by the barrier height. In particular, increasing $V_1$ enhances the effective isolation between the two subsystems, thereby modifying the tunneling rate and accelerating the dephasing-induced relaxation of the population imbalance.

Fig.~\ref{Fig3} provides further insight into the many-body dynamics underlying the weakly interacting regime, corresponding to the same parameters used in Fig.~\ref{Fig2}(b). Fig.~\ref{Fig3}(a) shows the time evolution of the occupations of the first two natural orbitals obtained from MCTDHB together with the corresponding occupations from the BH model. As the system evolves, the initially almost fully occupied first natural orbital decreases from $n_{1}\approx1$ to $n_{1}\approx0.5$, while the second orbital increases from $n_{2}\approx0$ to $n_{2}\approx0.5$, demonstrating the emergence of two-fold fragmentation. Although small oscillatory modulations become visible at longer times, these features are present in both MCTDHB and BH, indicating that they are intrinsic to the many-body dynamics rather than numerical artifacts. The excellent agreement between the BH and MCTDHB occupations over the entire evolution demonstrates that, in the weakly interacting regime, the dynamics is effectively described by two dominant natural orbitals. Consequently, the two-mode approximation captures the essential physics of fragmentation and Josephson dynamics in this regime.

Fig.~\ref{Fig3}(b)--(d) show the coefficient entropy $S_C(t)$, the orbital entropy $S_n(t)$, and the participation ratio (PR) as functions of time on a logarithmic scale, allowing both the short- and long-time dynamics to be clearly visualized. The orbital entropy $S_n(t)$ rapidly approaches a nearly constant value, reflecting the establishment of two-fold fragmentation. Likewise, $S_C(t)$ and the participation ratio evolve toward long-time quasi-stationary values, although their instantaneous values continue to exhibit finite fluctuations, as expected for a finite, closed quantum system undergoing unitary evolution. These fluctuations do not preclude equilibration; rather, they reflect finite-size effects about a coarse-grained equilibrium. Thus, in the weakly interacting regime, the many-body dynamics evolves toward a quasi-stationary fragmented state while retaining the intrinsic temporal fluctuations characteristic of isolated quantum systems.

\begin{figure}[t]
\centering
\includegraphics[width=0.45\textwidth]{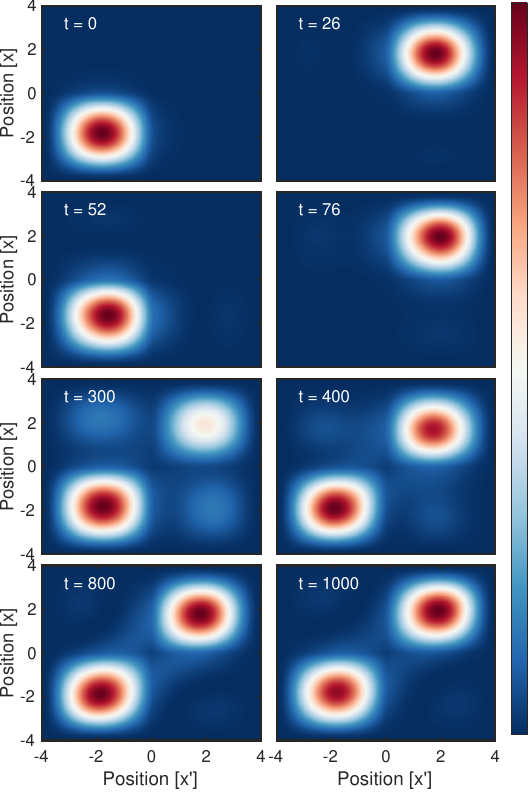}
\caption{Time evolution of the one-body density matrix $\rho^{(1)}(x, x^{\prime};t)$ following the same protocol as in Fig.~\ref{Fig2}(b). The initial state shows a lobe localized in the left reservoir due to population imbalance. As time evolves, tunneling leads to a redistribution of density, and at the fragmentation time two symmetric lobes emerge, corresponding to equal population in both wells. This balanced configuration is subsequently maintained during the long-time dynamics.}
\label{Fig4}
\end{figure}

The one-body reduced density matrix $\rho^{(1)}(x,x';t)$ provides further insight into the spatial redistribution of density during the dynamics, as shown in Fig.~\ref{Fig4}. Initially, $\rho^{(1)}(x,x';t)$ exhibits a pronounced lobe localized in the left reservoir, reflecting the prepared population imbalance. As the system evolves, atoms tunnel toward the right reservoir, and the density correspondingly redistributes across the junction.

At the characteristic decay time, when two-fold fragmentation is established, $\rho^{(1)}(x,x';t)$ develops two lobes of nearly equal weight, each localized in one of the reservoirs, indicating a symmetric redistribution of one-body coherence. This structure persists at long times, consistent with the equilibration of the population imbalance and the nearly equal occupations of the two leading natural orbitals. The evolution of $\rho^{(1)}(x,x';t)$ therefore provides a real-space manifestation of the crossover from an initially localized condensate to a fragmented state.

Overall, the weak-interaction regime is characterized by coherent Josephson dynamics with only weak many-body dephasing and gradual two-fold fragmentation. The excellent agreement between the BH and MCTDHB results demonstrates that the dynamics is effectively governed by two dominant modes throughout this regime. Consequently, the weak-interaction results establish a benchmark for the effective two-mode description and provide a well-defined reference against which the emergence of genuine multiorbital correlations, many-body relaxation, and the eventual breakdown of the BH model at stronger interactions can be assessed.

\subsection{Intermediate interaction regime: Josephson oscillations, many-body dephasing and relaxation}

We now turn to the intermediate interaction regime ($\Lambda=0.5$), where the interplay between interactions and initial population imbalance gives rise to qualitatively richer many-body dynamics. For the parameters considered here, the corresponding interaction parameters are $\gamma=0.4$ and $\zeta \approx 8.94$. The value of $\gamma$ indicates that interaction effects become relevant compared with the 1D density scale, while $\zeta>1$ implies that the healing length is smaller than the system size, allowing interaction-induced density variations and correlations to develop across the junction. This regime therefore represents the crossover from coherent Josephson dynamics, where the evolution is well described by an effective two-mode picture, toward a correlated many-body regime in which multiorbital effects, dephasing, and relaxation become increasingly important. In this regime, the comparison between MCTDHB and the BH model allows us to identify the onset of deviations from the effective two-mode description and to determine how multiorbital correlations contribute to the damping and relaxation of Josephson dynamics.

A key observation in this regime is that, for a fixed interaction strength, the dynamical behavior is predominantly controlled by the initial population imbalance, while the dependence on the barrier height $V_1$ becomes comparatively weak. Accordingly, throughout this section, the simulations are performed with $V_1=3.0$ in order to highlight the distinct dynamical features.

As the imbalance is increased, the system exhibits a clear sequence of dynamical responses that can be broadly classified into three regimes: (i) small imbalance, where coherent Josephson oscillations persist with weak fragmentation; (ii) intermediate imbalance, where many-body dephasing leads to collapse-and-revival dynamics; and (iii) large imbalance, where the system evolves toward a quasi-stationary state accompanied by significant fragmentation and saturation of many-body observables.

\begin{figure}[t]
\centering
\includegraphics[width=0.45\textwidth]{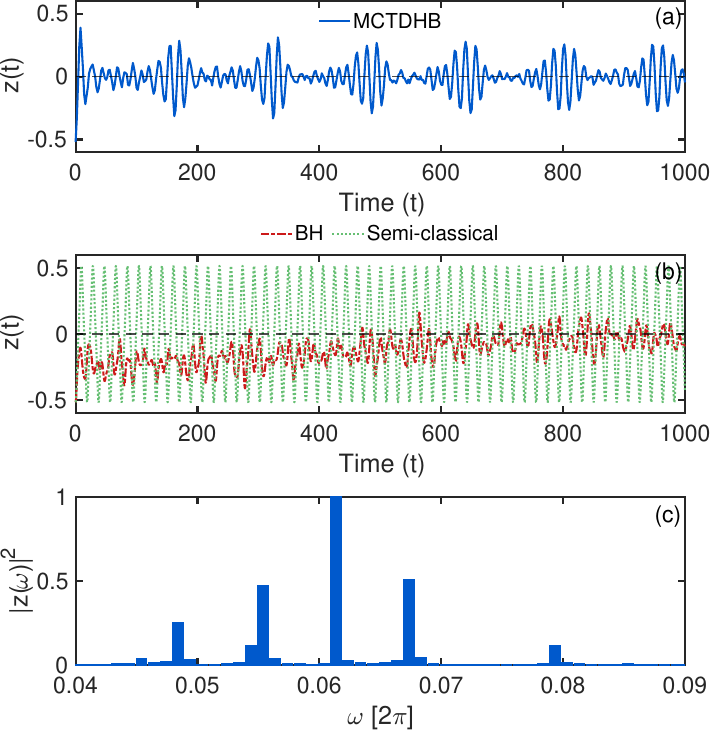}
\caption{Intermediate-imbalance dynamics in the intermediate interaction regime ($\Lambda=0.5$) for $V_1=3.0$ and $V_{\rm imb}=0.8$. (a) Time evolution of the population imbalance $z(t)$ obtained from MCTDHB, showing pronounced collapse-and-revival cycles due to many-body dephasing. (b) Population imbalance dynamics obtained from the full quantum BH model and its semiclassical two-mode approximation. (c) Frequency spectrum extracted from the MCTDHB dynamics, revealing multiple characteristic frequencies associated with the many-body dephasing process.}
\label{Fig5}
\end{figure}

We first consider the {\emph {small-imbalance regime}}, which serves as a reference point for the subsequent many-body dynamics. For small initial imbalances, the system exhibits nearly coherent Josephson oscillations with a well-defined dominant frequency. Although the interaction strength is larger than in the weakly interacting regime discussed previously, the dynamics remains close to a two-mode behavior, with only weak fragmentation and negligible population of higher natural orbitals. Accordingly, the many-body observables show only minor changes, indicating that correlation effects are still limited in this regime. Since the qualitative behavior closely resembles the coherent Josephson dynamics already discussed above, we focus below on the intermediate- and large-imbalance regimes, where many-body dephasing, fragmentation, and deviations from the two-mode description become prominent.

For \emph{intermediate initial imbalances}, the dynamics enters a regime where many-body dephasing becomes the dominant mechanism governing the evolution. Fig.~\ref{Fig5} presents the corresponding results for the intermediate interaction strength $\Lambda=0.5$, obtained for a fixed barrier height $V_1=3.0$ and imbalance offset parameter $V_{\rm imb}=0.8$. The corresponding BH parameters are $U=1.5915\times 10^{-1}$ and $J=7.2664 \times 10^{-2}$. Compared with the coherent Josephson oscillations observed at small imbalance, the dynamics exhibits pronounced correlation-induced modifications. The figure compares the population imbalance obtained from MCTDHB and the BH model, together with the corresponding oscillation-frequency analysis, thereby illustrating the onset of deviations from the effective two-mode description.

Fig.~\ref{Fig5}(a) shows the population imbalance $z(t)$ obtained from MCTDHB. In contrast to the nearly periodic Josephson oscillations observed for small imbalance, the oscillation amplitude now undergoes pronounced collapse-and-revival cycles over the evolution time. These revivals indicate that the apparent damping of the Josephson oscillations originates from many-body dephasing, where different many-body contributions evolve with distinct frequencies and subsequently rephase.

\begin{figure}[t]
\centering
\includegraphics[width=0.45\textwidth]{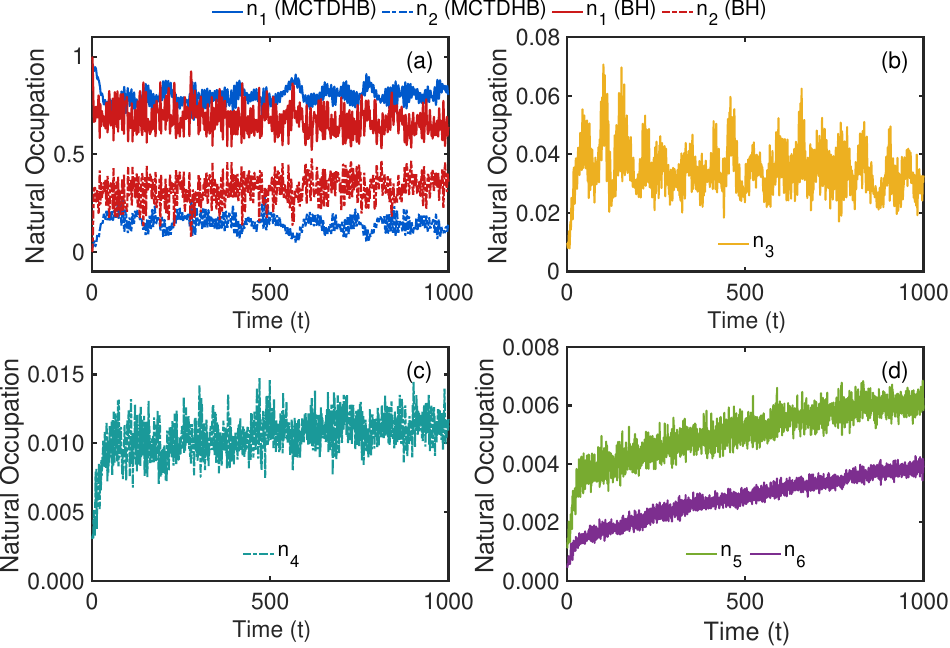}
\caption{Emergence of multiorbital effects in the intermediate-interaction regime following the same quench protocol as in Fig.~\ref{Fig5}. (a) Time evolution of the occupations of the two leading natural orbitals obtained from MCTDHB with $M=6$ orbitals compared with the corresponding occupations from the BH model. (b) Time evolution of the occupation of the third natural orbital $n_3(t)$, showing a finite population of approximately $0.02$--$0.04$. (c) Occupation of the fourth natural orbital $n_4(t)$, which remains smaller but contributes non-negligibly to the dynamics. (d) Occupations of the fifth and sixth natural orbitals, $n_5(t)$ and $n_6(t)$, which remain weakly populated with values of order $10^{-3}$. The finite population of higher natural orbitals demonstrates the emergence of multiorbital correlations and the gradual breakdown of the two-mode BH description.}
\label{Fig6}
\end{figure}

Fig.~\ref{Fig5}(b) presents the corresponding BH dynamics together with the semiclassical two-mode result. We observe that the interplay among the higher modes causes the population imbalance to relax to zero on a much shorter timescale, whereas within the two-mode BH approximation the relaxation extends over the entire duration of the simulation. This interpretation is supported by Fig.~\ref{Fig6}, which shows that the occupation of higher modes becomes significant during the dynamics. The semiclassical two-mode dynamics, as expected from its mean-field nature, shows persistent Josephson oscillations without damping and does not account for the dephasing and many-body effects present in the MCTDHB dynamics. These deviations indicate that the fixed two-mode description becomes insufficient in this regime, where additional orbitals and interaction-induced correlations contribute significantly to the dynamics.

Fig.~\ref{Fig5}(c) presents the frequency distribution extracted from the MCTDHB dynamics. The presence of multiple frequency components, rather than a single dominant Josephson frequency, reflects the contribution of different many-body modes to the evolution. The emergence of several characteristic frequencies provides direct evidence of many-body dephasing, where different components of the many-body wave function accumulate relative phases and subsequently interfere to produce the observed collapse-and-revival dynamics. Thus, the intermediate-interaction regime marks a departure from the simple two-mode Josephson picture and highlights the increasing importance of many-body correlations.

Fig.~\ref{Fig6}(a) compares the occupations of the leading natural orbitals obtained from MCTDHB ($M=6$) with the corresponding two-mode occupations from the BH model. In contrast to the weakly interacting regime, where the two approaches showed excellent agreement, clear deviations emerge in the present intermediate-interaction regime. The occupations of the two dominant MCTDHB natural orbitals no longer coincide with the BH predictions, indicating that the dynamics is not fully confined to an effective two-mode subspace. This difference arises because the MCTDHB orbitals are dynamically optimized and can redistribute occupation among multiple single-particle modes, whereas the BH model assumes a fixed two-mode basis throughout the evolution.

The origin of this deviation becomes evident from the occupations of the higher natural orbitals shown in Fig.~\ref{Fig6}(b). Although the first two orbitals remain dominant, additional orbitals acquire finite populations, with the third and fourth orbitals reaching occupations of approximately $n_3\sim0.02$--$0.04$ and $n_4\sim0.01$, respectively. The fifth and sixth orbitals, although weakly occupied with populations of order $10^{-3}$, also contribute to the many-body dynamics over long evolution times. These finite occupations demonstrate that the system develops multiorbital correlations beyond the two-mode description.

Therefore, the deviation between MCTDHB and BH in the intermediate regime is not due to a breakdown of the Josephson tunneling picture itself, but rather due to the emergence of additional correlated degrees of freedom that cannot be represented within a fixed two-mode framework. The population of higher natural orbitals provides direct evidence for the gradual breakdown of the BH approximation as interaction-induced correlations become increasingly important.

\begin{figure}[t]
\centering
\includegraphics[width=0.45\textwidth]{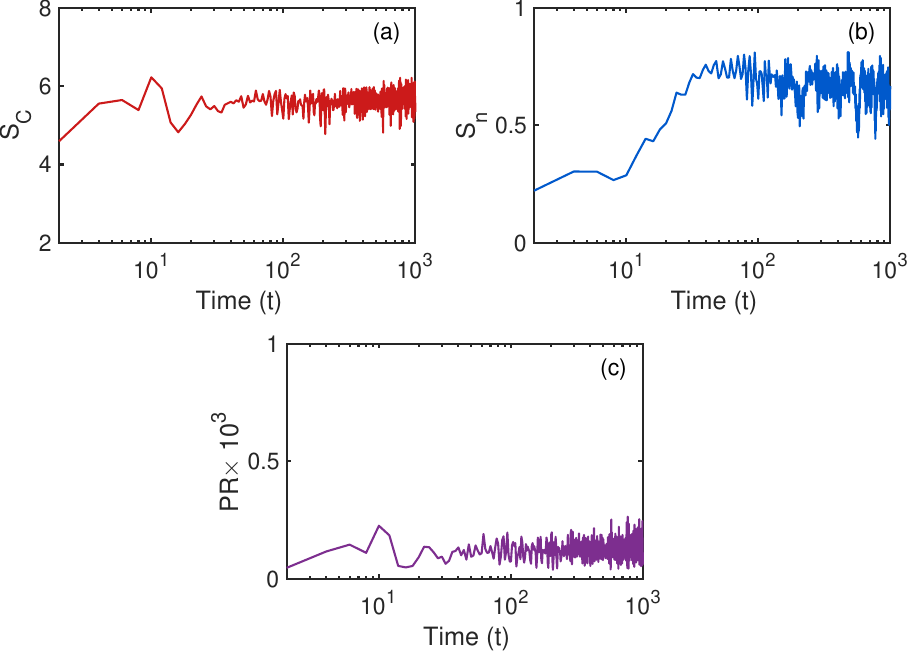}
\caption{Many-body dynamics in the intermediate-interaction regime following the same quench protocol as in Fig.~\ref{Fig5}. (a) Time evolution of the coefficient entropy $S_C(t)$, (b) orbital entropy $S_n(t)$, and (c) participation ratio (PR) obtained from MCTDHB using $M=6$ orbitals. The quantities are presented on a logarithmic time scale to resolve both the transient evolution and the long-time behavior. The entropy measures and PR reveal the development of fragmentation and the exploration of an extended Hilbert-space subspace beyond the effective two-mode description.}
\label{Fig7}
\end{figure}

Fig.~\ref{Fig7} provides further insight into the many-body dynamics through the evolution of the coefficient entropy $S_C(t)$, orbital entropy $S_n(t)$, and PR. The entropy measures and PR are presented on a logarithmic time scale, allowing both the transient evolution and the long-time behavior to be resolved. The orbital entropy $S_n(t)$ exhibits a relatively smooth increase followed by saturation, reflecting the development of fragmentation associated with the population of additional natural orbitals. In contrast, the coefficient entropy $S_C(t)$ and the PR show a initial growth followed by fluctuations around well-defined long-time averages. These finite fluctuations are expected for an isolated finite-size quantum system undergoing unitary evolution and do not preclude equilibration in the sense of coarse-grained observables. Instead, they indicate that the many-body wave function continues to explore a correlated region of Hilbert space while maintaining stable statistical properties at long times. Together with the occupation dynamics, these results demonstrate that the intermediate-interaction regime is characterized by many-body dephasing, fragmentation, and equilibration beyond the simple two-mode description.

Taken together, these results demonstrate that the intermediate-interaction regime represents the onset of genuinely many-body Josephson dynamics. While the BH model captures the initial tunneling behavior, it fails to reproduce the pronounced collapse-and-revival dynamics observed in the MCTDHB evolution. This discrepancy originates from the emergence of many-body correlations and the finite population of higher natural orbitals, which introduce additional dynamical frequencies beyond the effective two-mode description. The resulting multiorbital evolution leads to many-body dephasing and subsequent revival of the Josephson oscillations, manifested as collapse-and-revival behavior in the population imbalance. Thus, the intermediate regime provides a clear signature of the breakdown of the two-mode BH picture and highlights the necessity of a multiorbital many-body treatment to describe the correlated non-equilibrium dynamics. As the initial imbalance is further increased, these correlation effects become progressively stronger, leading beyond dephasing and revival dynamics toward the relaxation of coarse-grained observables and their long-time saturation, as discussed in the following section.

To further elucidate the underlying coherence properties, we analyze the corresponding one-body reduced density matrix. Its spatial structure and temporal evolution provide additional insight into the redistribution of coherence during the many-body dynamics and are presented in Appendix~\ref{appendix:a}.

\begin{figure}[t]
\centering
\includegraphics[width=0.45\textwidth]{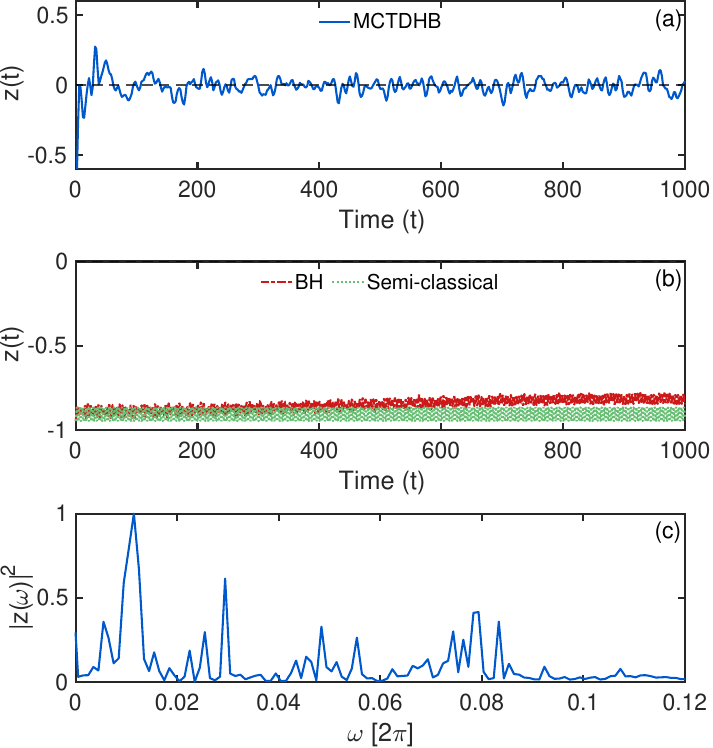}
\caption{Large-imbalance dynamics in the intermediate-interaction regime ($\Lambda=0.5$) for $V_1=3.0$ and $V_{\rm imb}=2.0$. (a) Population imbalance $z(t)$ obtained from MCTDHB, showing the rapid decay of Josephson oscillations and the subsequent relaxation toward a quasi-stationary state. (b) Comparison of the corresponding dynamics from the BH and semi-classical approaches, which remain close to the initial imbalance and fail to capture the relaxation observed in MCTDHB. (c) Frequency spectrum extracted from the MCTDHB dynamics, showing the presence of multiple characteristic frequencies associated with the many-body relaxation process.}
\label{Fig8}
\end{figure}

\begin{figure}[t]
\centering
\includegraphics[width=0.45\textwidth]{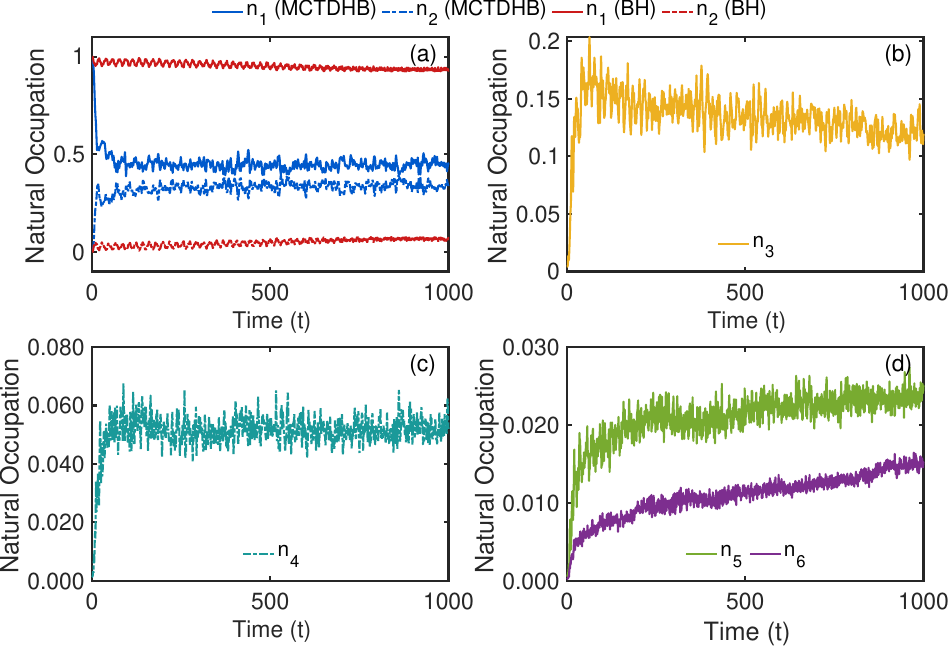}
\caption{Emergence of multiorbital effects in the large-imbalance regime following the same quench protocol as in Fig.~\ref{Fig8}. (a) Time evolution of the occupations of the two leading natural orbitals obtained from MCTDHB compared with the corresponding occupations from the BH model. The pronounced deviation between the two approaches indicates the breakdown of the effective two-mode description. (b) Occupation of the third natural orbital $n_3(t)$, showing a substantial population buildup during the dynamics. (c) Occupation of the fourth natural orbital $n_4(t)$, which acquires a comparable finite contribution. (d) Occupations of the higher natural orbitals $n_5(t)$ and $n_6(t)$, showing smaller but non-negligible contributions. The significant population of higher natural orbitals demonstrates the emergence of multiorbital correlations responsible for the relaxation dynamics beyond the BH description.}
\label{Fig9}
\end{figure}

For \emph{large initial imbalances}, the system enters a regime characterized by rapid redistribution of population and the emergence of relaxation behavior in coarse-grained observables. Fig.~\ref{Fig8} presents the dynamics in the large-imbalance regime for the intermediate interaction strength $\Lambda=0.5$. The barrier height is fixed at $V_1=3.0$, while the initial imbalance is increased to $V_{\rm imb}=2.0$. The corresponding BH parameters are the same as in Fig~\ref{Fig5}. In this regime, the dynamics changes qualitatively from the collapse-and-revival behavior observed at intermediate imbalance toward a relaxation process dominated by many-body correlations. The figure compares the population imbalance dynamics obtained from MCTDHB and the BH model, together with the corresponding frequency analysis, revealing the limitations of the effective two-mode description in capturing the relaxation dynamics.

Fig.~\ref{Fig8}(a) shows the population imbalance $z(t)$ obtained from MCTDHB. In contrast to the persistent oscillatory behavior observed for smaller imbalance, the initial Josephson oscillations rapidly decay, and $z(t)$ subsequently exhibits only small fluctuations around a value close to zero. This indicates an efficient redistribution of particles between the two reservoirs and the emergence of a quasi-stationary state in the population imbalance. The disappearance of coherent oscillations reflects the enhanced role of many-body correlations, which provide additional channels for dephasing and relaxation beyond the two-mode picture.

Fig.~\ref{Fig8}(b) compares the corresponding dynamics obtained from the BH and semi-classical approaches. Both curves remain trapped near the initial imbalance, with $z(t)$ showing no significant redistribution between the two reservoirs. Thus, neither the fixed two-mode BH model nor the semiclassical description captures the relaxation observed in MCTDHB. This demonstrates that, in the large-imbalance regime, the relaxation mechanism is intrinsically connected to many-body degrees of freedom and cannot be described within a reduced two-mode framework. Fig.~\ref{Fig8}(c) shows the frequency spectrum associated with the MCTDHB dynamics. Compared with the intermediate-imbalance regime, the spectrum contains a broader distribution of characteristic frequencies, reflecting the increased participation of many-body modes. The absence of a single dominant Josephson frequency is consistent with the rapid dephasing of the imbalance oscillations and the onset of relaxation dynamics.

Fig.~\ref{Fig9} provides further microscopic insight into the relaxation dynamics in the large-imbalance regime by analyzing the natural orbital occupations. Fig.~\ref{Fig9}(a) compares the occupations of the two leading natural orbitals obtained from MCTDHB with those from the BH model. In contrast to the intermediate-imbalance regime, where the deviations between the two approaches were moderate, the difference becomes pronounced in the present regime. The occupations predicted by MCTDHB no longer follow the BH dynamics, demonstrating that the system cannot be described within an effective two-mode framework.

The origin of this strong deviation is revealed by the population of higher natural orbitals shown in Figs.~\ref{Fig9}(b)--(d). The third and fourth natural orbitals acquire substantial occupations, reaching values of approximately $n_3\sim0.15$ and $n_4\sim0.06$, respectively. Such large populations indicate that multiple single-particle modes actively participate in the dynamics. Even higher orbitals, although less populated, contribute to the correlated many-body evolution. This strong multiorbital occupation provides a direct microscopic explanation for the relaxation observed in the population imbalance, as additional modes introduce new dynamical pathways for dephasing and redistribution of particles.

Thus, in the large-imbalance regime, the breakdown of the BH description is no longer a small quantitative correction but a qualitative failure of the two-mode approximation. The substantial occupation of higher natural orbitals demonstrates that the relaxation process involves genuine many-body correlations and additional dynamical pathways that are absent in the effective two-mode picture. As a consequence, the system explores an increasingly large portion of the many-body Hilbert space, motivating an analysis of statistical relaxation through entropy-based measures and the participation ratio.

\begin{figure}[htbp]
\centering
\includegraphics[width=0.45\textwidth]{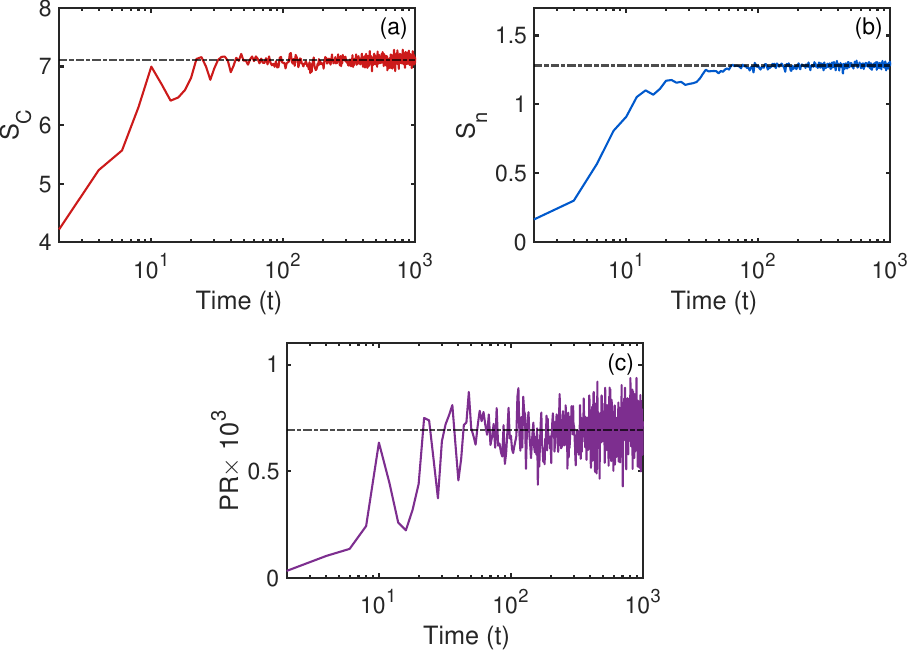}
\caption{Ergodicity analysis in the large-imbalance regime following the same quench protocol as in Fig.~\ref{Fig8}. Time evolution of (a) coefficient entropy $S_C(t)$, (b) orbital entropy $S_n(t)$, and (c) participation ratio (PR) obtained from MCTDHB. The corresponding long-time averaged values (black dashed dotted lines) are compared with the Gaussian orthogonal ensemble (GOE) predictions in the main text.
}
\label{Fig10}
\end{figure}

Fig.~\ref{Fig10} presents the time evolution of the many-body indicators $S_C(t)$, $S_n(t)$, and $\rm PR$, together with their corresponding long-time averaged values. To characterize the extent of Hilbert-space exploration and the nature of the relaxation dynamics, we compare these quantities with the predictions of the Gaussian orthogonal ensemble (GOE) of random matrices~\cite{Jaynes-I,Jaynes-II,Kota,Rigol,Izrailev_2012}. For a random matrix of dimension $D\times D$, the GOE estimates are $S_C^{\mathrm{GOE}}=\ln(0.48D)$ and ${\rm PR}^{\mathrm{GOE}}=D/3$~\cite{Kota}. Here, $D$ is taken as the number of many-body configurations $N_{\mathrm{conf}}=\binom{N+M-1}{N}$ that effectively participate in the dynamics. For the orbital entropy, the corresponding GOE reference value is determined by the number of contributing orbitals $M$, giving $S_n^{\mathrm{GOE}}=-\sum_{i=1}^{M}(1/M)\ln(1/M)=\ln M$. These GOE values provide quantitative benchmarks for assessing the extent to which the dynamics explores the available many-body Hilbert space.

For $N=10$ bosons and $M=6$ orbitals, the Hilbert-space configuration dimension is $N_{\mathrm{conf}}=3003$, yielding the GOE estimates $S_C^{\mathrm{GOE}}=\ln(0.48\times3003)=7.273$, ${\rm PR}^{\mathrm{GOE}}=1001$, and $S_n^{\mathrm{GOE}}=1.79$. The corresponding long-time averaged values obtained from our simulations are $S_C^{\mathrm{sat}}=7.109$, ${\rm PR}^{\mathrm{sat}}\approx 694$, and $S_n^{\mathrm{sat}}=1.282$. The close agreement of the coefficient entropy and participation ratio with their GOE values indicates that the many-body wave function explores a large fraction of the accessible Hilbert space, providing evidence for ergodic-like relaxation within the accessible many-body manifold. The remaining deviations originate from the finite interaction strength and finite system size, which restrict the accessible configurations compared with the ideal GOE limit. Furthermore, the lower value of $S_n^{\mathrm{sat}}$ compared with the GOE prediction indicates that the system develops strong but incomplete fragmentation, with the natural orbital occupations remaining non-uniform.

While the instantaneous quantities display finite temporal fluctuations, their long-time averages provide a more appropriate characterization of the relaxation properties of this finite system. The agreement between these coarse-grained quantities and the GOE benchmarks demonstrates ergodic-like relaxation within the accessible many-body Hilbert space rather than complete stationarity of the microscopic state.

To further characterize the irregular dynamics in the large-imbalance regime and to assess the sensitivity of the evolution to nearby initial conditions, we analyze the divergence of neighboring trajectories. Specifically, two simulations are initialized with slightly different population imbalances $V_{\rm imb}=2.0$ and $V_{\rm imb}=2.1$, and the separation between the corresponding trajectories is quantified by
\begin{equation}
d_z(t)=\left|z_1(t)-z_2(t)\right|,
\end{equation}
together with the corresponding effective finite-time Lyapunov exponent,
\begin{equation}
\lambda_{\mathrm{eff}}(t>0)=\frac{1}{t}\ln\left[\frac{d_z(t)}{d_z(0)}\right].
\end{equation}

\begin{figure}[htbp]
\centering
\includegraphics[width=0.45\textwidth]{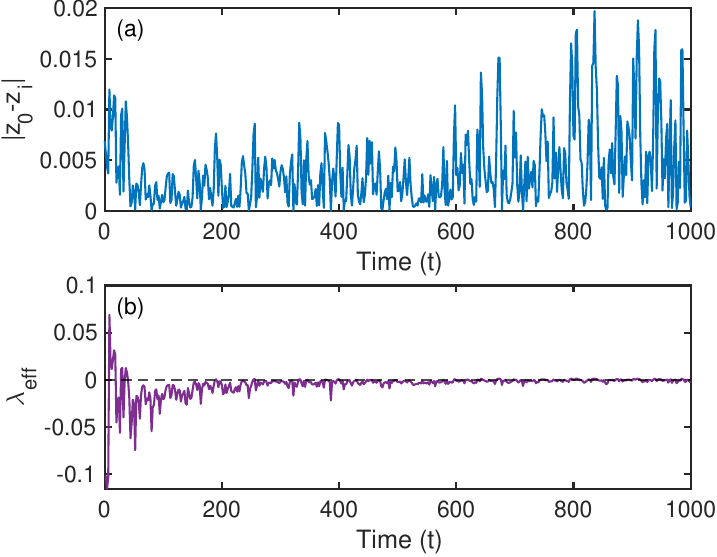}
\caption{Sensitivity to initial conditions in the large-imbalance regime for $V_1=3.0$. (a) Time evolution of the trajectory separation $d_z(t)=|z_1(t)-z_2(t)|$ for two simulations initialized with slightly different population imbalances $V_{\rm imb}=2.0$ and $V_{\rm imb}=2.1$. (b) Corresponding effective finite-time Lyapunov exponent $\lambda_{\mathrm{eff}}(t)$. The bounded trajectory separation and the decay of $\lambda_{\mathrm{eff}}(t)$ toward zero indicate the absence of sustained exponential divergence, consistent with finite quantum many-body dynamics.}
\label{Fig11}
\end{figure}

The results are presented in Fig.~\ref{Fig11}. Fig.~\ref{Fig11}(a) shows that the trajectory separation exhibits irregular temporal fluctuations but remains bounded throughout the evolution, without any sustained exponential growth. Consistently, the effective Lyapunov exponent shown in Fig.~\ref{Fig11}(b) does not converge to a positive constant and instead gradually approaches zero at long times. This behavior indicates the absence of a classical Lyapunov instability in the present finite quantum many-body system, where the dynamics evolves in a finite-dimensional Hilbert space rather than in a classical phase space.

Therefore, trajectory divergence alone does not provide a suitable criterion for identifying ergodic behavior in the present system. Instead, the ergodic character is inferred from the statistical properties of the many-body wave function. The approach of the long-time averaged coefficient entropy and participation ratio toward their corresponding GOE benchmarks demonstrates that the dynamics explores a large fraction of the accessible Hilbert space, while the finite orbital entropy indicates strong but incomplete fragmentation. Together, these complementary diagnostics provide evidence for ergodic-like relaxation in the accessible many-body Hilbert space. These results therefore support an interpretation of the long-time dynamics in terms of ergodic-like many-body relaxation, rather than classical chaos, within the finite Hilbert space of the present system. This interpretation is further corroborated by the evolution of the one-body reduced density matrix, discussed in Appendix~\ref{appendix:a}, which reveals the accompanying redistribution of one-body coherence during the relaxation process.

\subsection{Strong interaction regime: interaction-induced dynamical freezing}

We now turn to the strong-interaction regime ($\Lambda=5.0$), where the dynamics undergoes a qualitative transition from many-body relaxation to interaction-induced dynamical freezing. For the parameters considered here, the corresponding interaction parameters are $\gamma=4.0$ and $\zeta \approx 28.3$, indicating that the system lies in a strongly correlated regime where interaction effects dominate the dynamics. In this regime, strong repulsive interactions substantially suppress tunneling between the two reservoirs, leading to the breakdown of the Josephson dynamics observed at weaker interactions.

Unlike the intermediate-interaction regime, where increasing the initial imbalance promotes many-body dephasing and ergodic-like relaxation, the dominant effect of strong interactions is to inhibit particle transport and suppress large-scale collective motion. In this regime, the frozen dynamics originates from strong interaction-induced correlations, which cause the bosons to spatially avoid one another and develop well-separated density maxima. The resulting redistribution of the density strongly suppresses tunneling between the two reservoirs, leading to a pronounced reduction of particle transport and only small temporal evolution of the many-body observables. Consequently, this regime represents a qualitatively distinct non-equilibrium state, where the dynamics is governed primarily by strong interaction-induced correlations rather than by the dephasing and statistical relaxation mechanisms characteristic of the intermediate-interaction regime.

Since the dynamics in this regime is governed by strong multiorbital correlations, the effective two-mode BH description is no longer expected to provide a quantitatively reliable description. Accordingly, in the following we focus exclusively on the MCTDHB results.

To highlight the distinctive features of this regime, we consider two representative cases: small imbalance ($V_{\rm imb}=0.03$) with an intermediate barrier ($V_1=1.5$), and large imbalance ($V_{\rm imb}=2.0$) with a higher barrier ($V_1=3.0$). We further explore intermediate parameter choices that interpolate between these two limiting cases.

\begin{figure}[htbp]
\centering
\includegraphics[width=0.45\textwidth]{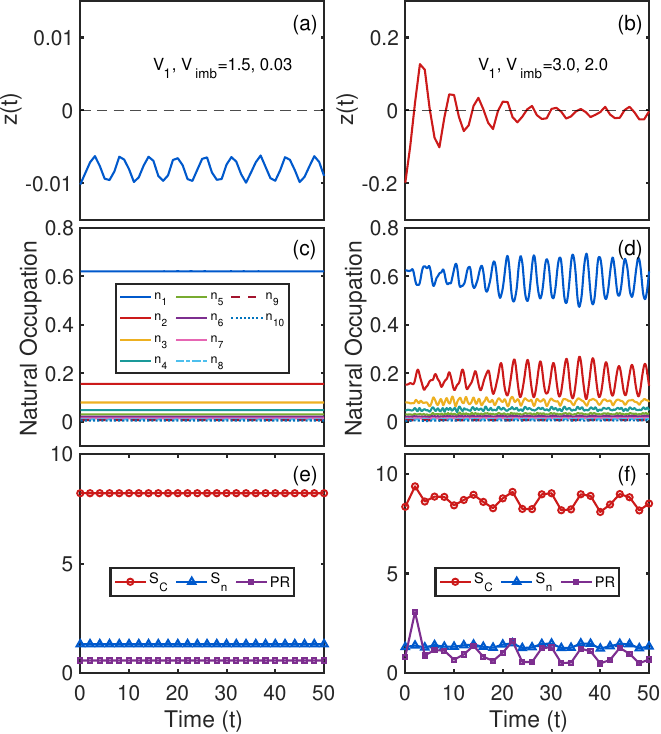}
\caption{(a),(b) Time evolution of the population imbalance $z(t)$ in the strong-interaction regime ($\Lambda=5.0$) for small imbalance $V_{\rm imb}=0.03$
(a) and large imbalance $V_{\rm imb}=2.0$ (b). (c),(d) corresponding natural occupations for small (c) and large (d) imbalance. (e), (f) represents coefficient entropy $S_{C}$, orbital entropy $S_n$ and participation ratio $\rm PR$ (scaled) for small imbalance $V_{\rm imb}=0.03$ and large imbalance $V_{\rm imb}=2.0$.  For small imbalance, $z(t)$
remains nearly frozen with negligible tunneling, while large imbalance exhibits a few initial oscillations followed by rapid decay. The dynamics reveal persistent fragmentation in both cases, with weak oscillatory behavior in the natural occupations for larger imbalance. Computation is done with $M=10$ orbitals.
}
\label{Fig12}
\end{figure}

The population imbalance $z(t)$ for small and large initial imbalances is shown in Figs.~\ref{Fig12}(a) and \ref{Fig12}(b), respectively. For small imbalance, $z(t)$ remains nearly frozen, exhibiting only small-amplitude oscillations about its initial value, indicating that tunneling between the two reservoirs is strongly suppressed. For large imbalance, the system undergoes only a few initial tunneling oscillations before the imbalance rapidly relaxes toward values close to zero, after which only weak residual fluctuations remain. This behavior demonstrates that strong interactions effectively suppress long-time coherent transport across the junction.

To accurately capture the correlated dynamics in this regime, we employ $M=10$ orbitals to ensure numerical convergence. The corresponding natural orbital occupations are presented in Figs.~\ref{Fig12}(c) and \ref{Fig12}(d). For small imbalance, several natural orbitals possess significant occupations, demonstrating pronounced many-body fragmentation even for a small initial imbalance. Moreover, the occupations remain nearly constant throughout the evolution, consistent with the frozen population dynamics. For large imbalance, the initial tunneling cycles are accompanied by weak oscillatory modulations of the natural occupations, reflecting a limited redistribution of particles among the occupied orbitals before the dynamics settles into the frozen regime.

This behavior is further reflected in the evolution of the coefficient entropy $S_C$, the orbital entropy $S_n$, and the participation ratio (PR), shown in Figs.~\ref{Fig12}(e) and \ref{Fig12}(f). For small imbalance, all three quantities remain nearly constant, consistent with the frozen population imbalance and the essentially time-independent fragmentation pattern. For large imbalance, $S_C$ and PR exhibit small but persistent oscillatory modulations that remain visible throughout the simulated time window (see Appendix~\ref{appendix:b}). These residual oscillations originate from coherent beating among a discrete set of many-body eigenstates populated during the quench and are therefore expected in a finite, isolated quantum system. They coexist with the suppression of long-time transport, indicating that interaction-induced dynamical freezing is accompanied by residual many-body coherence rather than complete microscopic stationarity.

\begin{figure}[htbp]
\centering
\includegraphics[width=0.45\textwidth]{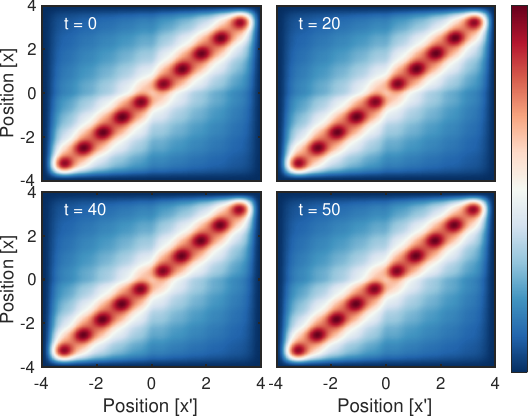}
\caption{Time evolution of the one-body density matrix $\rho^{(1)}(x,x^{\prime};t)$ in the strong-interaction regime ($\Lambda=5.0$) for small imbalance $V_{\rm imb}=0.03$, shown at $t=0,20,40,$ and $50$. The initial state exhibits ten well-separated lobes corresponding to the particles, reflecting strong interaction-induced correlations and suppressed spatial overlap. The structure remains essentially unchanged throughout the evolution, demonstrating interaction-induced dynamical freezing and strongly suppressed tunneling, consistent with the nearly constant population imbalance.}
\label{Fig13}
\end{figure}

\begin{figure}[htbp]
\centering
\includegraphics[width=0.45\textwidth]{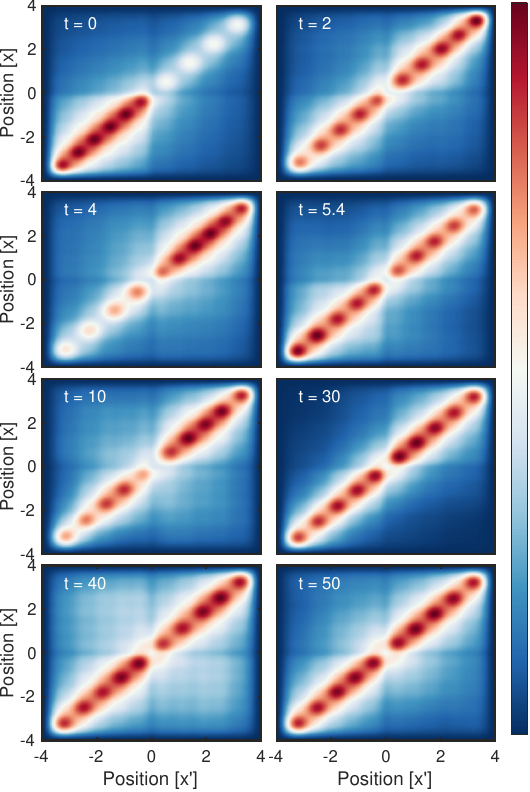}
\caption{Time evolution of the one-body density matrix  $\rho^{(1)}(x, x^{\prime};t)$, in the strong interaction regime ($\Lambda=5.0$) for large imbalance $V_{\rm imb}=2.0$,
shown at representative times. Initially ($t=0$), the two reservoirs exhibit unequal population and fragmentation, with stronger localization in the more populated well. Following the quench, particles redistribute, leading to alternating population and fragmentation between the wells. Around $t=30$, both reservoirs reach a balanced configuration, consistent with the equilibration of $z(t)$, which persists at longer times}
\label{Fig14}
\end{figure}

The one-body density matrix $\rho^{(1)}(x, x^{\prime};t)$,  shown in Fig.~\ref{Fig13}, provides clear evidence of frozen dynamics in the strongly interacting regime for small imbalance. At $t=0$, $\rho^{(1)}(x,x';t)$ exhibits ten well-defined bright lobes, corresponding to the ten particles, spatially separated and distributed across the two reservoirs. This structure reflects the strong interaction-induced correlations that suppress spatial overlap between the particles and significantly inhibit tunneling across the central barrier.

At later times ($t=20,40,50$), the structure of $\rho^{(1)}(x,x';t)$ remains essentially unchanged. The persistence of the well-separated density maxima indicates that the particles maintain their spatial separation throughout the evolution and that the many-body state remains dynamically frozen. This behavior is fully consistent with the nearly constant population imbalance $z(t)$, which exhibits only minimal oscillations, confirming the strong suppression of tunneling and the stability of the interaction-induced correlated state over time.

In the case of large imbalance, the time evolution of the one-body density matrix $\rho^{(1)}(x, x^{\prime};t)$ (Fig.~\ref{Fig14}) reveals a rich dynamical redistribution of both population and correlations. At $t=0$, the large imbalance leads to markedly different fragmentation in the two reservoirs: the left well, containing six particles, exhibits well-separated and strongly localized lobes, indicative of pronounced fragmentation, while the right well, with four particles, shows comparatively weaker fragmentation.

Immediately after the quench, when the two reservoirs are brought to equal depth, the particles begin to redistribute between the wells. This results in a gradual equalization of both population and fragmentation. At later times, the system exhibits a competing behavior: as one reservoir becomes more populated, the other correspondingly depletes, leading to alternating dominance in population and degree of fragmentation.

Around $t=30$, when the population imbalance $z(t)$ approaches equilibrium, both reservoirs display similar density structures, signaling the onset of equilibration. This balanced configuration persists at longer times, indicating that both population and correlations have reached a steady state.

In summary, the strongly interacting regime is characterized by a pronounced suppression of coherent tunneling and the emergence of interaction-induced dynamical freezing. For small imbalance, the system remains dynamically frozen, exhibiting negligible temporal evolution of both the density and the correlation properties. In contrast, for large imbalance, an initial redistribution of particles gives rise to transient dynamics, followed by the relaxation of both the population imbalance and fragmentation toward their long-time average values. Furthermore, for intermediate values of the barrier height and imbalance parameter (not shown here), the system exhibits irregular Josephson oscillations, indicating the breakdown of simple coherent dynamics in favor of strongly correlated, nonperiodic many-body evolution. Overall, these results demonstrate that strong interaction-induced correlations dominate the nonequilibrium dynamics, driving the system from coherent tunneling to regimes characterized by dynamical freezing, fragmentation, and suppressed particle transport.

To further substantiate the observed frozen dynamics in the strongly interacting regime, we present the long-time evolution in Appendix \ref{appendix:b}. The results confirm that the system remains dynamically constrained, with the relevant observables exhibiting negligible temporal variation even at extended times. {color{blue}A separate subsection of Appendix  \ref{appendix:b} presents a comparison with the dynamics of non-interacting fermions, providing a benchmark against the fermionized limit and further reinforcing the interpretation of the frozen dynamics observed in the strongly interacting regime.
}

\section{Conclusion}\label{sec:4}

In this work, we have presented a systematic many-body investigation of the non-equilibrium dynamics of a 1D BJJ confined in a box potential following an imbalance quench. By combining fully correlated MCTDHB simulations with systematic comparisons to the quantum BH model and its semiclassical counterpart, we have established the regime of validity of the effective two-mode description across weak, intermediate, and strong interaction regimes.

Our results demonstrate that the BH model accurately reproduces the coherent Josephson dynamics in the weakly interacting regime, where tunneling between the two reservoirs dominates the dynamics. As the interaction strength and the initial population imbalance increase, however, the agreement progressively deteriorates. The MCTDHB simulations reveal the dynamical occupation of higher natural orbitals, the growth of fragmentation, and the buildup of many-body correlations that are absent within the fixed two-mode description. These multiorbital processes drive the breakdown of the BH approximation and give rise to interaction-induced dephasing, relaxation, and eventually the suppression of coherent transport.

Beyond benchmarking the BH model, our many-body analysis provides a unified physical picture of the nonequilibrium dynamics. We show that fragmentation evolves from a static indicator of correlations into a dynamical mechanism governing the transition between coherent Josephson oscillations, many-body dephasing, ergodic-like relaxation, and interaction-induced dynamical freezing. The saturation of entropy measures and participation ratios toward their random-matrix benchmarks demonstrates that the intermediate-interaction regime is characterized by the exploration of an increasingly large fraction of the accessible many-body Hilbert space, whereas strong interactions suppress tunneling and constrain the dynamics to a frozen state.

Overall, this work provides a unified picture of the emergence and competition between coherent tunneling, many-body dephasing, equilibration, and dynamical freezing, while systematically delineating the regime of validity of the BH description for 1D box-confined Bose-Josephson junctions. Our results demonstrate that a fully correlated multiorbital treatment is essential for describing non-equilibrium transport beyond the weakly interacting regime, where higher-orbital occupations and many-body correlations fundamentally reshape the Josephson dynamics.

Several important questions remain open. In particular, it will be important to understand how the dynamical generation of fragmentation and the associated relaxation mechanisms evolve with particle number, system size, and dimensionality, and how they are are modified by long-range interactions, different confinement geometries, and disorder. Addressing these questions will provide further insight into the role of many-body correlations in nonequilibrium transport and equilibration in quantum fluids.

\section{Acknowledgments}
A.~K.~S. acknowledges support from JSPS KAKENHI Grant Number JP25K00925. F.~C. acknowledges Fundação de Amparo à Pesquisa do Estado de São Paulo(FAPESP), Grant No. 2023/17459-8. A.~G. acknowledges support from Fundação de Amparo à Pesquisa do Estado de São Paulo (FAPESP) [Grant 2024/01533-7], and Conselho Nacional de Desenvolvimento Científico e Tecnológico (CNPq) [Grant 306219/2022-0]. R.~D. acknowledges support from the French government under the France 2030 investment plan, as part of the Initiative d’Excellence d’Aix-Marseille Université – AMIDEX AMX-22-CEI-069.

\appendix

\section{MCTDHX units}\label{appendix:c}
We introduce a characteristic length $\ell$ and solve the Schr\"odinger equation in dimensionless form. The Hamiltonian is rescaled by $\hbar^2 / (m\ell^2)$, which defines the natural energy scale of the system as $E=\hbar^2/m\ell^2$. Correspondingly, the natural time scale is given by $t=m\ell^2 / \hbar$. In typical experiments, the system size is on the order of
$\mathcal{O}(10~\mu \text{m})$~\cite{PhysRevLett.110.200406,PhysRevLett.95.010402,Navon2015}. To relate the dimensionless quantities to experimental parameters, for example, consider $^{87}$Rb atoms and choose $\ell \approx 10~\mu \text{m}$, consistent with typical trapping lengths. This yields the characteristic energy and time scales $E=\hbar^2/m\ell^2\approx 7.7\times 10^{-34}$ J and $t=m\ell^2 / \hbar \approx 0.137$ s, respectively.

\section{Semi-classical Josephson equations}\label{appendix:semiclassical}

For completeness, we briefly summarize the semi-classical description of the BJJ, which is obtained from the two-site BH model in the mean-field (large-$N$) limit. In this regime, the system dynamics can be expressed in terms of the population imbalance $z(t) = (n_L - n_R)/N$ and the relative phase $\phi(t)$ between the two modes.

Within this approximation, the equations of motion take the form
\begin{eqnarray}
\dot{z} &=& -2J \sqrt{1 - z^2}\,\sin\phi,\\
\dot{\phi} &=& UN z + 2J \frac{z}{\sqrt{1 - z^2}} \cos\phi,
\end{eqnarray}
where $J$ denotes the tunneling amplitude and $U$ the interaction strength.

These equations describe the Josephson oscillations in the mean-field limit and correspond to a classical Hamiltonian dynamics in the conjugate variables $(z,\phi)$. In particular, they provide a useful reference for understanding coherent oscillatory behavior in the weakly interacting regime.

However, this description neglects quantum fluctuations and many-body correlations beyond mean-field theory. As a result, it cannot capture phenomena such as many-body dephasing, equilibration, or fragmentation-induced dynamical effects that emerge in the full quantum dynamics. These effects are instead described within the MCTDHB framework used in the main text.

\begin{figure}[t]
\centering
\includegraphics[width=0.45\textwidth]{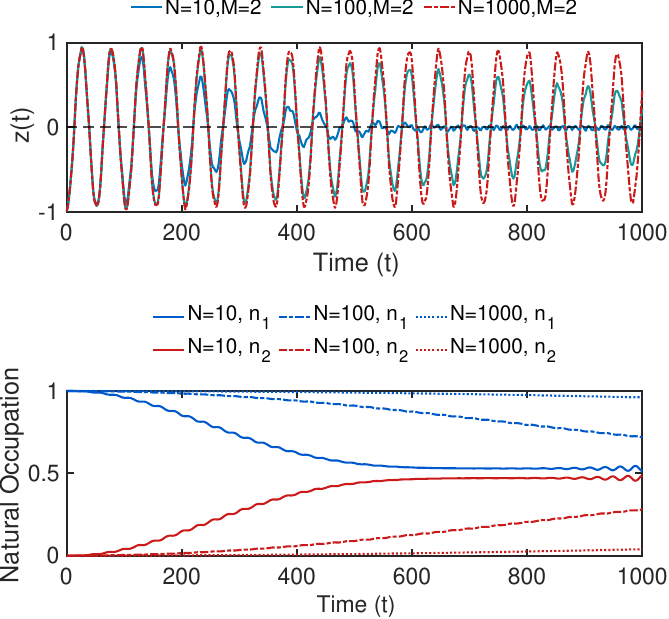}
\caption{Particle-number scaling of the many-body dynamics under the mean-field interaction scaling protocol. (a) Population imbalance $z(t)$ for different particle numbers $N=10$, $100$, and $1000$ obtained with $M=2$ orbitals. Increasing the particle number delays the collapse of the Josephson oscillations and extends the timescale of coherent dynamics. (b) Corresponding occupations of the two leading natural orbitals, $n_1(t)$ and $n_2(t)$. For $N=10$, the occupations gradually approach a two-fold fragmented state, while for larger particle numbers the redistribution of orbital populations occurs on much longer timescales.}
\label{Fig15}
\end{figure}

\section{Finite-size and mean-field interaction scaling}\label{appendix:finitesize}

To investigate the possible extension of our results toward larger particle numbers, we perform an additional finite-size scaling analysis following the standard mean-field interaction scaling protocol $\Lambda=\Lambda_0(N-1)$. We consider systems with $N=10$, $100$, and $1000$ particles and adjust the interaction parameter $\Lambda_0$ such that the corresponding mean-field interaction strength remains comparable. In the Gross-Pitaevskii mean-field theory, systems with different particle numbers but the same value of the mean-field parameter exhibit identical dynamics. This scaling procedure allows us to examine the role of finite-$N$ effects and assess how the many-body dynamics evolves toward the large particle-number regime.

The calculations are performed with $M=2$ orbitals. Fig.~\ref{Fig15} presents the particle-number scaling of the dynamics, with the population imbalance $z(t)$ shown in panel (a) and the occupations of the two leading natural orbitals, $n_1(t)$ and $n_2(t)$, shown in panel (b), following the same quench protocol as in Fig.~\ref{Fig2}(b). For $N=10$, the system exhibits a rapid collapse of the Josephson oscillations in Fig.~\ref{Fig15}(a), accompanied by a gradual redistribution of the natural orbital occupations in Fig.~\ref{Fig15}(b). In particular, the initial dominant occupation of the first natural orbital decreases, while the second orbital gains population, leading to approximately equal occupations of the two leading orbitals at long times. This two-fold fragmentation is consistent with the collapse of the population imbalance and demonstrates the connection between the loss of coherent tunneling dynamics and the redistribution of population among the natural orbitals. The agreement between the $M=2$ and $M=4$ calculations for $N=10$ (Fig.~\ref{Fig2}(b), Fig.~\ref{Fig3}(a)) confirms the convergence of the orbital description within this scaling analysis.

With increasing particle number, both the collapse of the population imbalance and the redistribution of natural orbital occupations occur on progressively longer timescales. For $N=100$, the oscillation amplitude of $z(t)$ decreases in Fig.~\ref{Fig15}(a), but a complete collapse is not reached within the simulated time window up to $t=1000$. Correspondingly, Fig.~\ref{Fig15}(b) shows that the leading natural orbital remains predominantly occupied, and significant two-fold fragmentation does not develop within the accessible evolution time. For $N=1000$, only weak damping of the imbalance oscillations is visible, while $n_1(t)$ and $n_2(t)$ remain close to their initial values over the same time interval. Nevertheless, the observed trend indicates that the fragmentation process is not eliminated with increasing particle number, but is instead shifted toward much longer evolution times. It is worth noting that, for $N=1000$, the dynamics shown in Fig.~\ref{Fig15}(a) closely matches the semiclassical dynamics presented in Fig.~\ref{Fig2}(b). This comparison explicitly demonstrates that the semiclassical description becomes increasingly accurate as the particle number $N$ increases.

\section{Evolution of coherence from the one-body reduced density matrix with $\Lambda=0.5$}\label{appendix:a}

\subsection{Intermediate imbalance regime}
\begin{figure}[t]
\centering
\includegraphics[width=0.45\textwidth]{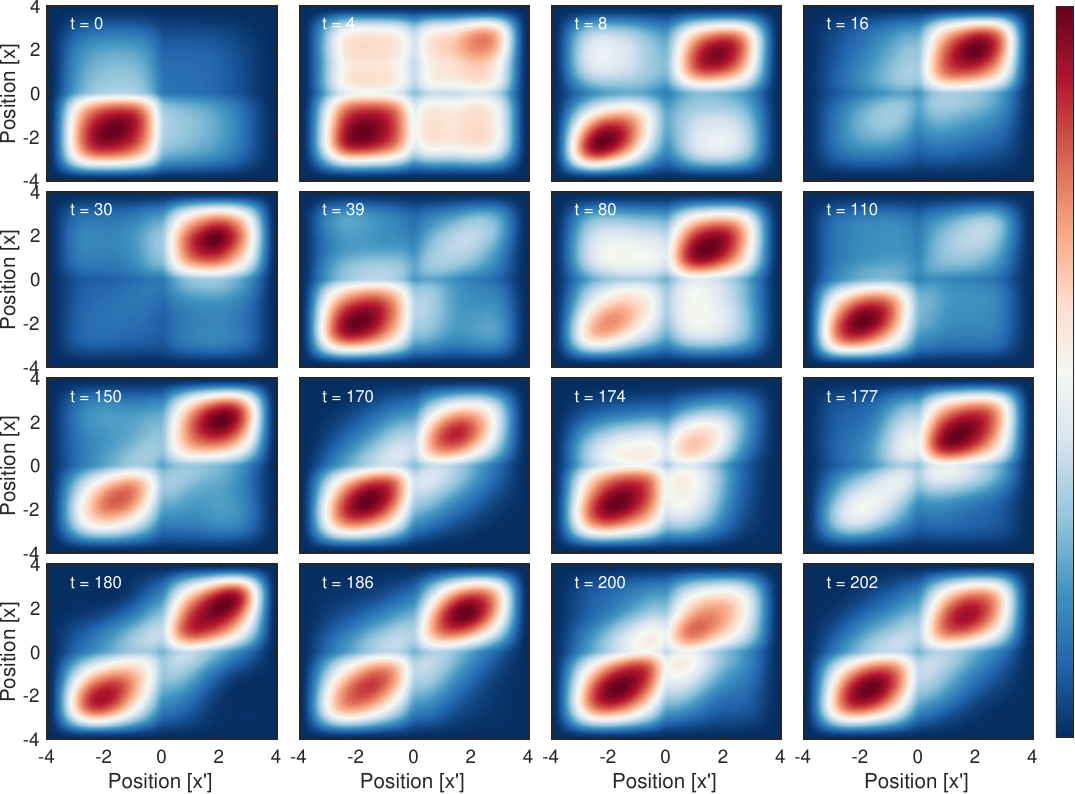}
\caption{Time evolution of the one-body reduced density matrix $\rho^{(1)}(x, x^{\prime};t)$, in the intermediate interaction regime ($\Lambda=0.5$) for a intermediate imbalance $V_{\rm imb}=0.8$,
and a barrier height $V_1=3.0$. The dynamics illustrate the redistribution of coherence across the junction, characterized by the decay and revival of correlations associated with dephasing and rephasing processes.}
\label{Fig16}
\end{figure}

To clearly demonstrate a complete cycle of dephasing and rephasing, we analyze the dynamics at 16 representative time points (Fig.~\ref{Fig16}). The evolution reflects a correlated flow of coherence across the junction. Initially, the system is localized in the left well with a high degree of phase coherence. As tunneling sets in, coherence is progressively redistributed between the two wells, leading to dephasing manifested by the spreading of correlations over multiple many-body modes. This results in a transient equilibration regime, where the population appears balanced but coherence is significantly reduced. At later times, partial rephasing occurs as correlations reorganize, driving the system toward localization in the right well. The subsequent return of the density to the left well is accompanied by further rephasing. Thus, the spatial motion of the density is directly tied to the temporal redistribution of correlations, with collapse and revival emerging from the interplay of dephasing and rephasing processes.

\begin{figure}[t]
\centering
\includegraphics[width=0.45\textwidth]{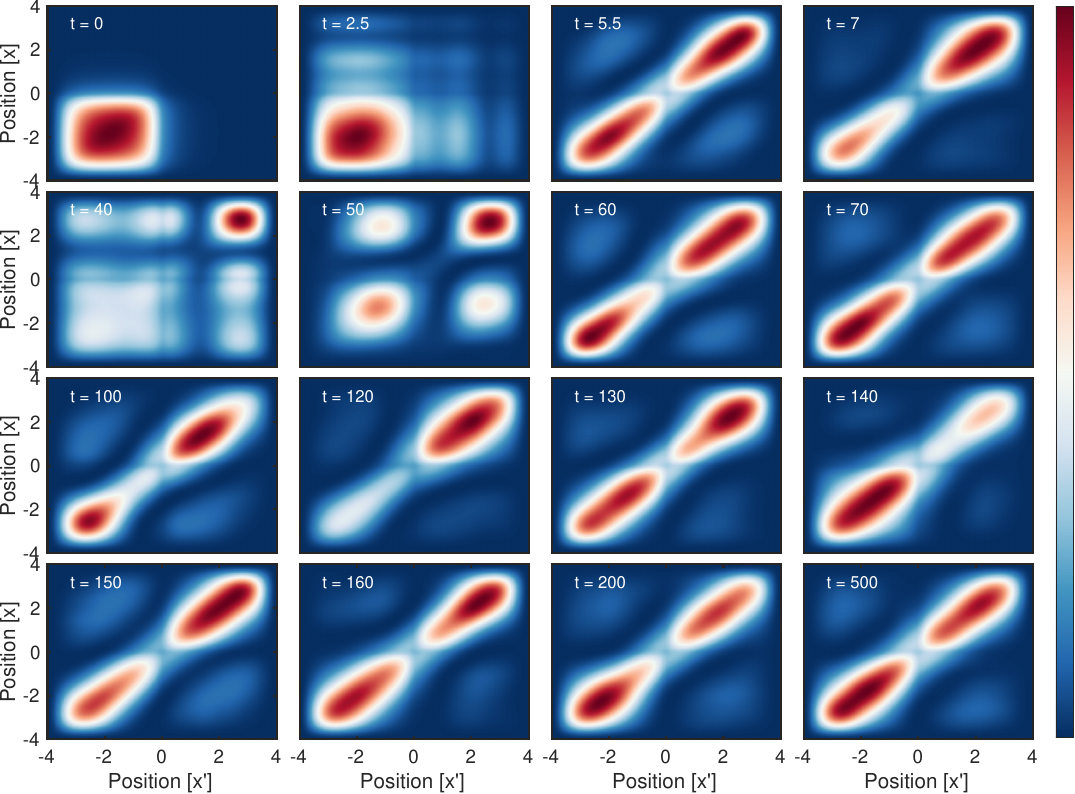}
\caption{Time evolution of the one-body reduced density matrix $\rho^{(1)}(x, x^{\prime};t)$, in the intermediate interaction regime ($\Lambda=0.5$) for a large imbalance $V_{\rm imb}=2.0$, and a barrier height $V_1=3.0$, shown at 16 representative time points from early to long times. The system is initially localized in the left well, followed by tunneling toward the right well and subsequent equilibration. After only a few tunneling events, the system approaches a near-stationary configuration, indicating fast relaxation, in contrast to the collapse–revival dynamics observed at intermediate imbalance.
}
\label{Fig17}
\end{figure}

\subsection{Large imbalance regime}
 To capture the evolution of coherence across different timescales, we analyze the one-body reduced density matrix at 16 representative time points, spanning from early-time dynamics to the long-time limit (Fig.~\ref{Fig17}). In the large-imbalance regime, the system is initially highly localized in the left well, reflecting the large population asymmetry imposed at $t=0$. At short times, tunneling from the left to the right well sets in; however, unlike the intermediate imbalance case, this transfer does not lead to sustained oscillatory or cyclic behavior. Instead, after only a few tunneling events, the system rapidly approaches to an equilibrated configuration with no subsequent rephasing to restore the initial state. This behavior stands in sharp contrast to the intermediate regime, where coherence undergoes repeated cycles of dephasing and rephasing, giving rise to collapse and revival dynamics. Here, the absence of revival signal genuine many-body relaxation, with the system effectively settling into a steady state at long times.

\begin{figure}[t]
\centering
\includegraphics[width=0.45\textwidth]{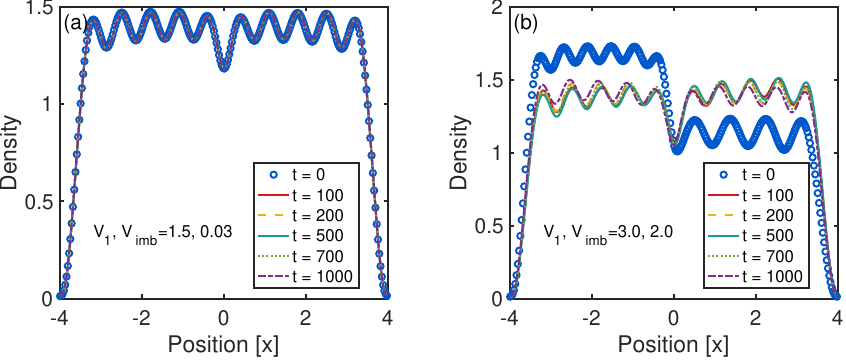}
\caption{Time evolution of the one-body density $\rho(x,t)$ over the entire long-time dynamics. (a) For small imbalance $V_{\rm imb}=0.03$ and the central barrier height $V_1=1.5$. (b) For large imbalance $V_{\rm imb}=2.0$ the central barrier height $V_1=3.0$. The small-imbalance case exhibits fully frozen behavior with no visible modulation, while for large imbalance, weak temporal variations persist due to asymmetric fragmentation, although the system remains in a highly correlated regime.
}
\label{Fig18}
\end{figure}

\section{strongly interacting regime with $\Lambda=5.0$}\label{appendix:b}
\subsection{Long-time dynamics}
\begin{figure}[t]
\centering
\includegraphics[width=0.45\textwidth]{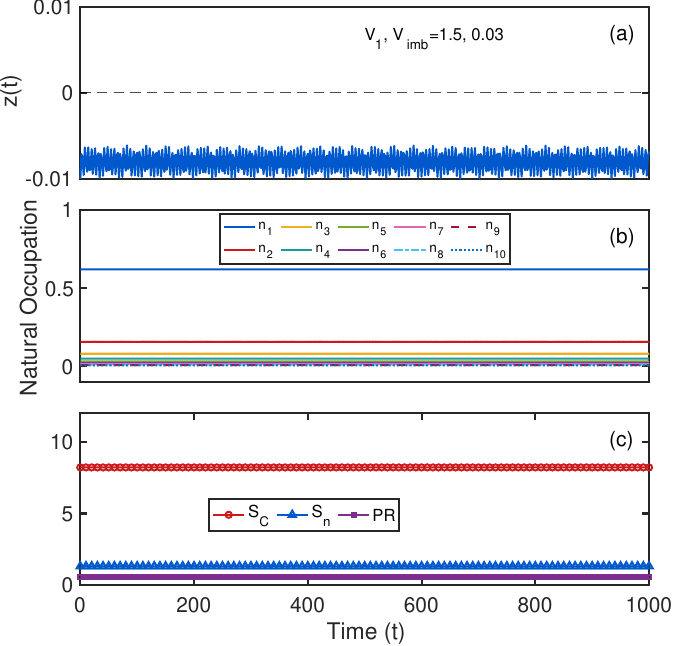}
\caption{(a) Long-time evolution of the population imbalance $z(t)$ for small imbalance $V_{\rm imb}=0.03$ and barrier strength $V_1=1.5$. (b) Corresponding dynamics of the natural orbital occupations. (c) Time evolution of the coefficient entropy $S_C$, orbital entropy $S_n$,  and participation ratio (PR) (scaled) for the same parameters. All observables exhibit frozen dynamics with negligible temporal variation.}
\label{Fig19}
\end{figure}

\begin{figure}[pb]
\centering
\includegraphics[width=0.4\textwidth]{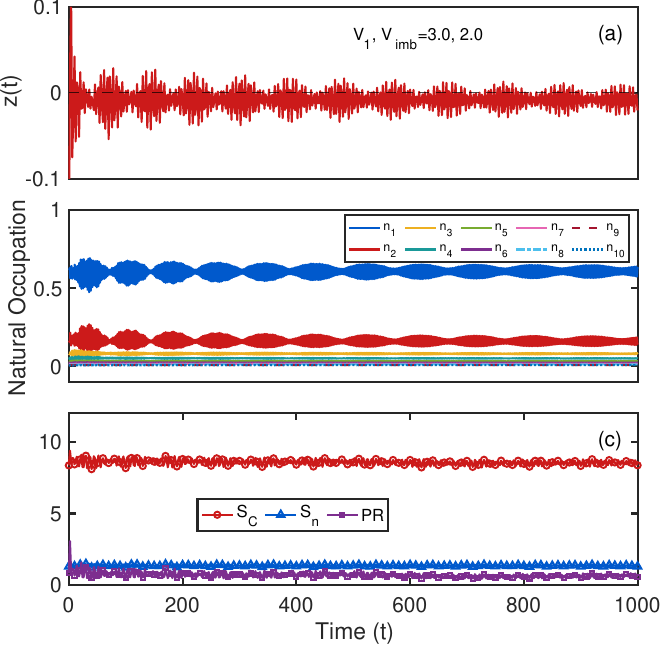}
\caption{(a) Long-time evolution of the population imbalance $z(t)$ for large initial imbalance $V_{\rm imb}=2.0$ and barrier strength $V_1=3.0$. The dynamics exhibit closely spaced oscillatory packets, indicating rapid dephasing and rephasing processes that are not individually resolved. (b) Corresponding natural orbital occupations. (c) Time evolution of the coefficient entropy $S_C$, orbital entropy $S_n$, and participation ratio (PR) (scaled). All observables display similar fast, weak modulations, reflecting the compressed timescales of the underlying many-body dynamics.
}
\label{Fig20}
\end{figure}

In this section, we examine the long-time dynamics through the evolution of the one-body density $\rho(x,t)$, the population imbalance $z(t)$, and the entropy measures, as shown in Figs.~\ref{Fig18}--\ref{Fig20}. Fig.~\ref{Fig18} presents $\rho(x,t)$ at representative times $t=0,100,200,500,700,$ and $1000$ for small and large initial imbalances.

For the small-imbalance case (Fig.~\ref{Fig18}(a)), the density profile remains essentially unchanged throughout the evolution, indicating robust interaction-induced dynamical freezing and strongly suppressed particle transport between the two wells. The fragmentation pattern also remains stable, demonstrating that both the density distribution and the underlying many-body correlations are preserved over long times.

For the large-imbalance case (Fig.~\ref{Fig18}(b)), the density exhibits weak but persistent temporal modulations arising from the asymmetric redistribution of particles between the two wells. Nevertheless, the density remains highly structured throughout the evolution, indicating that the system retains strong interaction-induced correlations despite the residual dynamics.

Fig.~\ref{Fig19} shows the long-time evolution of the population imbalance $z(t)$, the natural orbital occupations, and the entropy measures for the small-imbalance case ($V_{\rm imb}=0.03$). All observables remain nearly constant throughout the evolution. The population imbalance exhibits only negligible oscillations, confirming the strong suppression of tunneling, while the natural occupations show no appreciable redistribution among the orbitals, demonstrating that the fragmentation pattern is preserved. Likewise, the coefficient entropy $S_C$, occupation entropy $S_n$, and participation ratio $\rm PR$ remain essentially unchanged, indicating that the many-body state explores no additional configurations during the evolution. Together, these results demonstrate the remarkable stability of the dynamically frozen state over long times.

We next consider the large-imbalance case ($V_{\rm imb}=2.0$), shown in Fig.~\ref{Fig20}. The population imbalance $z(t)$ exhibits a sequence of closely spaced oscillatory packets, indicating that the underlying dephasing and rephasing processes occur on much shorter timescales than in the intermediate-interaction regime. Consequently, individual collapse-and-revival cycles are no longer well resolved, and the dynamics appear as nearly continuous oscillatory packets. Similar behavior is observed in the natural orbital occupations and the entropy measures, which display weak, rapidly varying modulations. These results demonstrate that, although coherent many-body dynamics persist, the characteristic timescales are substantially compressed by the strong interactions, producing a qualitatively different long-time evolution from that of the intermediate-interaction regime.

\begin{figure}[htbp]
\centering
\includegraphics[width=0.4\textwidth]{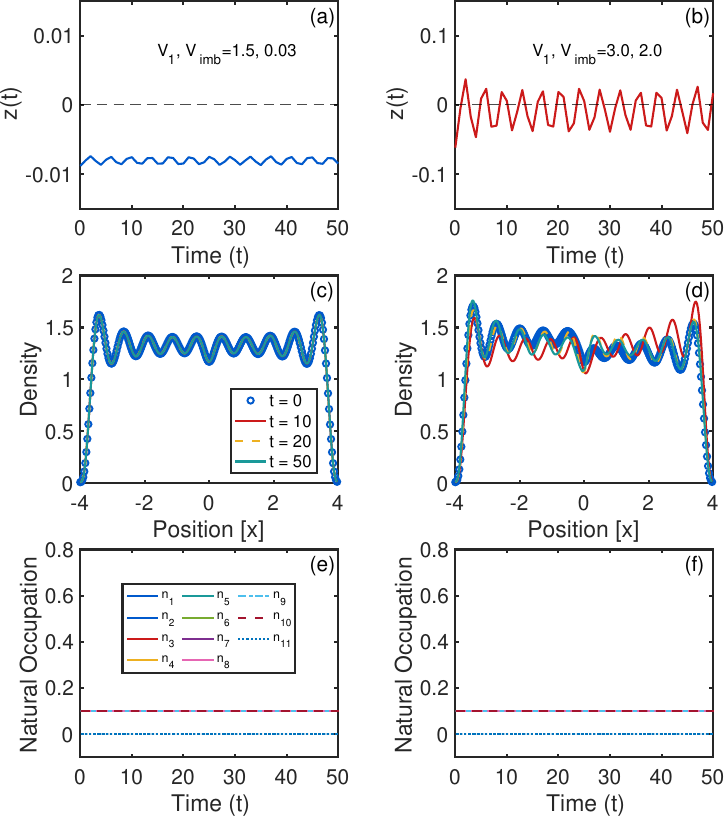}
\caption{Dynamics of $N=10$ non-interacting fermions for two representative box-trap configurations: $(V_1,V_{\rm imb})=(1.5,0.03)$ (left column) and $(V_1,V_{\rm imb})=(3.0,2.0)$ (right column). Panels (a) and (b) show the population imbalance, $z(t)$; panels (c) and (d) depict the evolution of the one-body density, $\rho^{(1)}(x,t)$; and panels (e) and (f) present the occupations of the natural orbitals. All results are shown up to $t=50$. The fermionic system exhibits dynamical freezing for the lower barrier and smaller imbalance, while for the higher barrier and larger imbalance a pronounced temporal modulation of the density due to particle redistribution between the two wells is observed. Unlike the strongly interacting bosonic case, the non-interacting fermions remain completely ten-fold fragmented throughout the dynamics, consistent with the Pauli exclusion principle.}
\label{Fig21}
\end{figure}

\subsection{Comparison with Non-Interacting Fermions}

Although the strongest interaction considered in this work corresponds to a large interaction parameter, $\gamma = 4.0$ and $\zeta = 28.3$, the system remains well below the Tonks--Girardeau (TG) limit, which is attained only for $\gamma \gg 1$. Nevertheless, it is instructive to compare the many-body dynamics with those of non-interacting fermions, to which impenetrable bosons can be mapped in the TG regime. Such a comparison provides a useful benchmark for assessing the degree of fermionization in our system and demonstrates that, despite the strong interactions, the observed dynamics remain quantitatively distinct from those of the fermionized limit.

Fig.~\ref{Fig21} presents a comparison between the dynamics of the strongly interacting Bose gas and those of ten non-interacting fermions in the same box-trap potential and consider the same quench protocol computed with the MCTDH-X software~\cite{lin:2020,MCTDHX}. For bosons, the symmetrized many-body configuration corresponds to a permanent, while for fermions the antisymmetrized state is a Slater determinant. Due to the Pauli principle, fermions require at least $M \ge  N+1 $ orbitals. We consider the same two representative cases discussed in the main text: $(V_1,V_{\rm imb})=(1.5,0.03)$ and $(V_1,V_{\rm imb})=(3.0,2.0)$. For a direct comparison, all observables are presented up to $t=50$.

The population imbalance, $z(t)$, is shown in Fig.~\ref{Fig21}(a) and Fig.~\ref{Fig21}(b), which may be directly compared with the corresponding bosonic results presented in Fig.~\ref{Fig12}(a) and Fig.~\ref{Fig12}(b). For the lower barrier and smaller initial imbalance (Fig.~\ref{Fig21}(a)), the non-interacting fermions exhibit dynamical freezing in excellent qualitative and quantitative agreement with the strongly interacting bosonic system. In contrast, for the stronger barrier and larger initial imbalance (Fig.~\ref{Fig21}(b)), only qualitative agreement is observed. While the strongly interacting bosons undergo a few tunneling oscillations before relaxing to an almost vanishing population imbalance, the non-interacting fermions continue to exhibit weak oscillations about zero.

The corresponding one-body density is displayed in Fig.~\ref{Fig21}(c) and Fig.~\ref{Fig21}(d). For the lower barrier and smaller imbalance, the density consists of ten nearly equally distributed peaks throughout the evolution, reflecting the occupation of ten distinct single-particle states and closely resembling the qualitative structure shown in Fig.~\ref{Fig18}(a). In contrast, for the stronger barrier and larger initial imbalance, a pronounced temporal modulation of the density is observed due to the redistribution of particles between the two wells. The overall evolution is qualitatively similar to that shown in Fig.~\ref{Fig18}(b), although the details of the density distribution differ from those of the interacting bosonic system.

Finally, Fig.~\ref{Fig21}(e) and Fig.~\ref{Fig21}(f) present the occupations of the natural orbitals. To satisfy the Pauli exclusion principle, the calculations are performed using $M=11$ orbitals for $N=10$ non-interacting fermions. As expected, the first ten natural orbitals remain singly occupied throughout the evolution, corresponding to complete ten-fold fragmentation. This behavior is in sharp contrast to the strongly interacting bosonic results shown in Fig.~\ref{Fig12}(c) and Fig.~\ref{Fig12}(d), where only incomplete fragmentation is observed. These results demonstrate that, although the bosonic system considered in the main text lies in a strongly interacting regime, it remains quantitatively distinct from the fully fermionized Tonks--Girardeau limit.

\bibliography{biblio}

\end{document}